# Interpolation and Prewar-Postwar Output Volatility and Shock-Persistence Debate: A Closer Look and New Results*

**Hashem Dezhbakhsh**
Department of Economics, Emory University
Atlanta, GA 30322, USA
econhd@emory.edu

**Daniel Levy****
Department of Economics, Bar-Ilan University
Ramat-Gan 5290002, Israel
Department of Economics, Emory University
Atlanta, GA 30322, USA
ICEA, ISET at TSU, and RCEA
Daniel.Levy@biu.ac.il

February 10, 2026

It is well established that the U.S. prewar output was more volatile and less shock persistent than the postwar output. This is often attributed to the data interpolation employed to construct the prewar series. Our analytical results, however, indicate that commonly used linear interpolation has the opposite effect on shock persistence and volatility of a series—it increases shock persistence and reduces volatility. The surprising implication of this finding is that the actual differences between the volatility and shock persistence of the prewar and postwar output series are likely greater than the existing literature recognizes, and interpolation has dampened rather than magnified this difference. Consequently, the view that postwar output was more stable than prewar output because of the effectiveness of the postwar stabilization policies and institutional changes has considerable merit. Our results hold for parsimonious stationary and nonstationary time series commonly used to model macroeconomic time series.

* This is a significantly expanded version of a working paper we had circulated earlier with the title "The Role of Interpolation in the Prewar-Postwar Output Volatility Debate: A Closer Look and New Results." We thank Theodoros Bratis, Bob Chirinko, Caroline Fohlin, Philip Franses, Daniel Kaufmann, Essie Maasoumi, Juan Rubio Ramirez, Jesús Vázquez, and Tao Zha for helpful discussions and comments on the earlier draft and the participants of the 2025 Annual Conference of the Economic History Association of Israel and the 2025 Annual Conference of the Society for Economic Measurement for useful suggestions. Louis Johnston kindly answered our questions about the use of linear interpolation in constructing historical time series. Any remaining errors are ours.

** Corresponding author: Daniel Levy, Daniel.Levy@biu.ac.il

"The existing estimates of gross national product for the 70 years before World War II have done more to shape economists' perceptions of prewar business cycles than any other macroeconomic series. ….. Hence, much of what economists believe about prewar cyclical fluctuations is derived directly from the cyclical behavior of prewar GNP. As a result, the accuracy of the prewar estimates of GNP is one of the main determinants of the accuracy of our views about the prewar cycle."

**Christina Romer (1989, p. 1)**

"There are two possible explanations of this apparently different behavior [of GNP] before and after World War I. The first is that shocks to GNP have in fact been more persistent in the latter part of the twentieth century than they were in the nineteenth and early twentieth centuries. The second explanation is that the relatively low serial correlation is a spurious result of the interpolation procedure used to construct the annual estimates for the early period."

**James Stock and Mark Watson (1986, p. 149)**

## 1. Introduction

A large body of research by macroeconomists and economic historians is devoted to studying the difference between the characteristics of prewar (WWI) and postwar (WWII) output and other macroeconomic aggregates. Analyses of these differences contribute importantly to our understanding of the macroeconomy and business cycles as well as the role that stabilization policies and institutional changes play in macroeconomic fluctuations (Burns 1960, Lucas 1977, Romer 1989, Bergman et al. 1998, Basu and Taylor 1999). As Romer (1991) emphasizes, the two particularly important characteristics that these studies focus on are (i) the volatility and (ii) persistence of short-run movements in output and other macroeconomic aggregates.

Burns (1960), Lucas (1977), Bailey (1978), DeLong and Summers (1986, 1988), Tobin (1980), Backus and Kehoe (1992), Zarnowitz (1992, 1998), Fuhrer and Schuh (1998), Samuelson (1998), and Taylor (1986), among others, argue that the prewar US economy was far more volatile than the postwar economy. Stock and Watson (1986), DeLong and Summers (1986), and Campbell and Mankiw (1987a, 1987b) also report that prewar and postwar US aggregate time series differ in their shock persistence properties, with the prewar series exhibiting lower shock persistence than the postwar series.

Romer, however, challenges these views, stating (1986a, p. 314) that the prewar-postwar difference in output behavior is a figment of the data. She argues that much of what we know about the behavior of output, prices, employment, and more generally, business cycles before World War I, might have been influenced by the interpolation of historical time series (Romer 1986a, 1986b, 1989, 1991, 1992, 1994, and 1999). This argument draws on the fact that many historical macroeconomic data are observed at low or irregular frequencies or contain missing observations, therefore, to increase the frequency or to generate missing observations, linear interpolation has been commonly used because of its simplicity and intuitiveness. Interpolation is, therefore, viewed as one of the key factors that may have contributed to the differences found between the prewar and postwar business cycles.[1] Romer (1986c, p. 343, and 1989, p. 18) also argues that the use of linear interpolation to construct the missing prewar unemployment data is a possible source of the excess volatility found in Lebergott's (1964) labor force participation and unemployment series.

In this paper, we present analytical results that are contrary to these arguments. Specifically, we show that linear interpolation *increases* the shock persistence and *reduces* the volatility of both stationary and nonstationary series. Since it is not feasible to *empirically* examine the conjectured effect of interpolation on time series data, given the limited availability of non-interpolated prewar data, we explore these issues by analytically modeling linear interpolation of time series and deriving volatility and shock persistence measures for both original, non-interpolated, and interpolated time series, and comparing them. We study five parsimonious data-generating processes (DGPs) commonly used in modeling macroeconomic time series, including stationary and nonstationary (difference-stationary) processes.

Our findings are as follows. First, linear interpolation reduces the volatility of the series and increases its shock persistence in all models examined, regardless of whether the original series is a pure random walk, a random walk with ARMA (1,1) errors, or a stationary AR(1), MA(1), or ARMA(1,1) process. Second, the size of these changes depends on the underlying DGP and the length of the interpolated segment (number of missing data points). Third, for all interpolated series, the variance ratio measure of shock persistence is larger than the corresponding theoretical value of the original noninterpolated series. Fourth, variances of all interpolated series we derive exhibit periodicity, which is consistent with Romer's (1986a, 1986b, 1986c, 1986d, 1989) argument regarding the interpolation-induced cyclical patterns in prewar data.

These findings imply that the difference between the volatility and shock persistence of the U.S. prewar and postwar macroeconomic series reported in the literature is not necessarily the artifact of linear interpolation of prewar data, as suspected. If at all, linear interpolation (a) should have reduced the volatility of the output series, not increased it, and (b) should have increased shock persistence, not reduced it. To the extent that linear interpolation was employed to generate the prewar data series, and low-order, parsimonious time series DGPs can approximate them well, our findings lead to a surprising conclusion that the true differences in the shock persistence and volatility of the prewar and postwar series may be far larger than the existing literature recognizes. That is because without interpolation, the prewar output would have been more volatile and less shock persistent.

---

[1] For example, Stock and Watson (1986), Jaeger (1990), and Dezhbakhsh and Levy (2022) suggest that the prewar-postwar differences in macroeconomic time series properties may be due to the interpolation procedure used to construct the prewar series. Charles and Darné (2010) draw similar conclusions based on robust unit root tests and also point to linear interpolation as the probable cause. Dezhbakhsh and Levy (1994) show that interpolation introduces periodicity in the moments and the cross-moments of the series, further conjecturing that it might affect a series' shock persistence properties. Franses (2022) and Dezhbakhsh and Levy (1994) show that linear interpolation alters the autocorrelation structure of the time series, which, if ignored, can lead to spurious results.

The merit of our work is borne out by the importance of historical data in shaping our macroeconomic understanding and the widespread use of linear interpolation in data construction. Refuting the argument that interpolation drives the widely accepted stylized fact that the postwar output was less volatile and more shock-persistent than the prewar output, and presenting analytical results that show the opposite, suggests that the effectiveness of postwar stabilization policy and institutional changes may have indeed contributed importantly to prewar-postwar differences, as argued by Burns (1960) in his AEA Presidential address, as well as by others.

The paper is organized as follows. In section 2, we discuss linear interpolation, its widespread applications in economics and other fields, and its expected impact on time series and their properties. In section 3, we develop a framework for modeling linear interpolation and describe the measures of volatility and shock persistence we employ. In section 4, we derive the shock persistence properties of a stationary series with and without interpolation for a variety of parametric specifications. In section 5, we do the same for nonstationary series by deriving the shock persistence properties of a nonstationary series with and without interpolation. In section 6, we examine the effect of interpolation on volatility by comparing the short and long variances of the non-interpolated and corresponding interpolated series. In section 7, we discuss the macroeconomic implications of our findings in light of the prewar vs. postwar output volatility and shock persistence debate. In section 8, we offer concluding remarks and caveats.

**2. Interpolation and Economic Time Series**

Some economic time series, and particularly historical data, are available only at low frequency. Examples include prewar data in the US, Europe, and other countries that were not available on an annual basis, quarterly series extracted from annual data, or weekly retail data that contain missing observations. For example, the Census of Manufacturers Data—one of the most important historical series that also include GDP, GDP deflator, and unemployment rate—were collected, for example, once every 10 years during the period 1869–1899, once every 5 years during the period 1899–1914, and once every 2 years during the period 1914–1919 (Stock and Watson, 1986).

In such situations, linear interpolation is often used to increase the frequency of the available data. For example, much of the prewar data on output and other aggregate variables posted at www.MeasuringWorth.com, a popular historical data archive, was constructed using linear interpolation. Other examples include the widely used time series constructed by Kuznets (1946, 1961), Shaw (1947), and Friedman and Schwartz (1982). The calendar period averaging, which statistical agencies like the US Bureau of Labor Statistics often employ for data estimation, is also a form of interpolation. For example, monthly CPI data for the period January 1913–August 1940 were constructed by the US Bureau of Labor Statistics using calendar period averaging (BLS 1966, p. 10).[2]

The use of linear interpolation to generate missing observations is not limited to statistical agencies. For example, Lebergott (1964) constructed an annual labor force participation rate for the 1900-1939 period by linearly interpolating between census years and applying the interpolated rates to population data for the intercensal years. Other applications include Levy and Chen (1994) for constructing aggregate quarterly capital stock series from annual data, Ma and Chu (2014) for constructing China's GDP data for 1840–1912, Johnson and Williamson (2018) for constructing GDP series, and Levy et al. (2020) for generating the missing observations of weekly time series of retail price series for the 1989-1997 period. Karger and Wray (2024) use linearly interpolated income data to study the lifetime earnings gap between whites and blacks.

Moreover, many international historical series have a linearly interpolated component. Cogley (1990), for example, analyses historical data for 9 countries: Australia, Canada, Denmark, France, Italy, Norway, Sweden, the UK, and the US, where the source of these data, except for the US, is Madison's (1982).[3] In discussing his data, Maddison (1982, Appendix A) refers to country-specific sources for this data. Our examination of these sources reveals, perhaps not surprisingly, that various types of linear interpolation were used in building historical time series of output for most of these countries. Examples include linear interpolation of the output series themselves, or linear interpolation of the component series, or linear interpolation of the price indexes used in deflating the nominal output values, etc.

In historical studies of financial markets, observed asset prices are often treated as representative of the market, which poses a selection bias because of thin trading that is endemic to such data (Rousseau 2009, Campbell et al. 2018a and 2018b). For example, Case et al. (2025) show that on the Brussels Stock Exchange, on average, 42.5% of corporate bonds did not trade at all in a given year. To overcome this problem, linear interpolation is frequently employed to fill in for the missing data (see, e.g., Fohlin 2025).

---

[2] Linear interpolation is also widely used in other disciplines. In epidemiology, COVID-related death figures that were reported during the pandemic were mostly based on interpolation (Katz and Sanger-Katz, *New York Times*, March 28, 2020). In Archeology, Carleton et al. (2014) show that the reported effects of cyclical droughts of the 1st millennium on Classic Maya society are artifacts of the bias introduced through linear interpolation. Other data interpolation examples include Adorf (1995) in astronomy, Jane et al. (2016) in bioinformatics, Liu (2016) and Liu and Hauskrecht (2016) in medicine, and Shi (2015) and Shi et al. (2017) in forestry.
[3] For the US, Cogley uses data from Balke and Gordon (1986), who do not employ interpolation methods for estimating their output series. Some of the international data series that Backus and Kehoe (1992) employ are either interpolated or contain interpolated components.

The potential adverse impacts of interpolation on stochastic properties of series and inference drawn from them have been suggested by several authors. For example, Balke and Gordon (1989) and Romer (1989) discuss the extent to which interpolation might have tainted key macroeconomic findings. Kaufmann (2020) suggests that CPI interpolation may have caused a misclassification of the 19$^{th}$-century inflationary and deflationary periods as well as the underestimation of the shortfall in economic activity during deflationary periods. Charles et al. (2014) find that if one uses the interpolated real GDP series of Johnston and Williamson (2018), then more than 50% of the peaks and troughs in the US business cycles, as identified by the NBER and Davis (2006), are altered. Cheung and Chinn (1997) and Murray and Nelson (2000) also conjecture that linear interpolation may have altered the statistical properties of historical time series. In line with Dezhbakhsh and Levy (1994), Franses et al. (2006) and Franses (2013) argue that interpolation is a possible cause of the periodicity found in the variance of many macroeconomic series they analyze. Moreover, Johnston and Williamson (2003) find that interpolated historical US GDP series exhibit fragile short-term properties, while Ehrmann (2000) discusses the challenges that interpolated series pose when assessing the monetary policy transmission across EU countries.[4]

Several studies conjecture that interpolation might have changed the shock persistence properties of the prewar macroeconomic series (e.g., Stock and Watson 1986, Jaeger 1990, Dezhbakhsh and Levy 1994, and Charles and Darné 2010). Others have argued that interpolation might have been the cause of the lower shock persistence observed in the US prewar economic series (e.g., Stock and Watson 1986, Jaeger 1990, Dezhbakhsh and Levy 2022).[5]

In this paper, we go beyond criticism of linear interpolation through cautionary notes or conjectures about its likely impact. Instead, we analytically assess the effect of linear interpolation

---

[4] Several studies have also pointed out the drawbacks of using interpolation on methodological grounds. For example, Douglas (1930, p. 56) argues that the convenient assumption of even changes is not only logically improbable but also refutable given the uneven changes in observed data. Hanes (2006, p. 156) holds similar views about Long's (1960) CPI series constructed from scanty retail price and rent data. Keating and Valcarcel (2015) highlight the quality differences between the interpolated pre-1929 and actual post-1929 real GDP data. Johansson (1967) provides GDP estimates for Sweden for 1861–1955 but warns against using his data for studying Swedish business cycles because, in his view, short-term cyclical fluctuations are likely to be concealed by various types of interpolations and extrapolations used in constructing the GNP data.

[5] Jaeger's (1990) finding that interpolation reduces shock persistence is based on a simulation where he assumes that the interpolation involves a padding term that is identical to the moving average component of the assumed true DGP. Dezhbakhsh and Levy (2022) show that Jaeger's simulation results can be analytically verified. However, his assumption that interpolated series are constructed with full knowledge of the true DGP is unrealistic.

on the statistical properties of time series.

## 3. Framework for Analyzing Linear Interpolation

Consider a DGP denoted by $\{Y_T\}$ where $T$ denotes time. To model missing values and interpolation effect, we divide the series generated by this process into segments of equal length $s$, where only one observation is available in each segment, and the rest may be missing. Based on this segmentation, the series $\{Y_T\}$ can be denoted by $y_{t,i}$, where the two subscripts, respectively, identify the segment and the location of the period within a segment.

Accordingly, the series $Y_T$ can be rewritten as $y_{t,i}$, $i = 1, 2, ..., s$, $s > 1$ and $t = 1, 2, ...$, with the following equivalency:

(1) $Y_1 = y_{1,1}$, $Y_2 = y_{1,2}$, ..., $Y_{s-1} = y_{1,s-1}$, $Y_s = y_{1,s}$, $Y_{s+1} = y_{2,1}$,..., and, in general, $Y_{s(t-1)+i} = y_{t,i}$.

For example, for quarterly data, the subscripts for $y_{t,i}$ denote years $t$, and quarters $i$. Thus, a 10-year-long quarterly series will be denoted by $y_{t,i}$, $t = 1, 2, ..., 10$ and $i = 1, 2, 3, 4$, while the original series are $Y_1, Y_2, ..., Y_{40}$. In the context of historical settings cited in section 2, $t$ and $i$, respectively, denote decades and years within the decades, where $s = 10$, and $t$ denotes decades. This would fit the case of interpolating missing annual observations from decennial benchmark data, as it was often done by economic historians and macroeconomists.

More generally, however, if $t$ is an arbitrary time interval, then $i$ denotes equal, non-overlapping sub-periods within that time interval. Therefore, when the series contains no missing values, we observe $y_{t,i} \forall t, i$, but when there are missing values, we only observe $y_{t,i} \forall t \mid i = s$. Linear interpolation is used to obtain the missing values of the latter series $\forall t \mid i \neq s$, resulting in a new series which we denote by $x_{t,i}$, where $x_{t,i} = y_{t,i} \forall t \mid i = s$ for the benchmark observations.

As noted, many historical series that were not collected on a regular basis had to be constructed by means of linear interpolation of benchmark observations. The missing $s - 1$ observations were generated for each $t$ using a segmented linear interpolation. Following this practice, we define

(2) $x_{t,i} = \frac{i}{s}(y_{t,s} - y_{t-1,s}) + y_{t-1,s} = \frac{i}{s}y_{t,s} + \frac{s-i}{s}y_{t-1,s}$.

Note that in the benchmark periods, the original non-interpolated series $y_{t,i}$ and the corresponding interpolated series $x_{t,i}$ coincide by construction; e.g., $x_{t,i} = y_{t,i}$ when $i = s$.

To characterize shock persistence properties, or long memory, of a time series, we adopt

Cochrane's (1988) measure of shock persistence, which for the series $Y_T$ is given by $V = \sigma_k^2 / k\sigma_1^2$, where $\sigma_1^2 = \text{var}(Y_T - Y_{T-1})$ denotes the "short variance," i.e., the variance of the series' 1-period growth, and $\sigma_k^2 = \text{var}(Y_T - Y_{T-k})$ denotes the "long variance," i.e., the variance of the series' cumulative growth over $k$-periods. As Stock (1994, p. 2741) notes, "… the presence of a unit root in output implies that shocks to output have great persistence through base drift, which can even exceed the magnitude of the original shock if there is positive feedback in the form of positive autocorrelation" (Campbell and Mankiw 1987a, 1987b, Cochrane 1988). In other words, the variance ratio measure of Cochrane (1988) is intended to ascertain whether or not the series converges to previous forecasts following a shock or diverges from these forecasts (Campbell and Mankiw 1987a, 1987b). The variance ratio reflects the magnitude of the variance of the shocks to the permanent component of the series relative to the variance of yearly growth rates. A larger value reflects a longer memory and thus more persistence. We use the variance of 1-period output growth, $\sigma_1^2 = \text{var}(Y_T - Y_{T-1})$, as a measure of output volatility similar to Acemoglu et al (2003) and Ramey and Ramey (1995), for example.

We derive the variance ratio measure for (i) the original non-interpolated series $y_{t,i}$, and for (ii) the linearly interpolated series $x_{t,i}$, obtaining

$$(3) \quad V_y = \frac{\text{var}(y_{t,i} - y_{t-1,i})}{k \, \text{var}(y_{t,i} - y_{t,i-1})} = \frac{\sigma_{k,y}^2}{k\sigma_{1,y}^2}, \quad \text{and} \quad V_x = \frac{\text{var}(x_{t,i} - x_{t-1,i})}{k \, \text{var}(x_{t,i} - x_{t,i-1})} = \frac{\sigma_{k,x}^2}{k\sigma_{1,x}^2}$$

and assuming, without loss of generality, that the displacement lag $k$ equals $s$. The two measures are then compared to identify the effect of interpolation on the persistence properties of the interpolated series.

In our analysis, we consider both stationary and nonstationary series. For stationary cases, we use AR(1), MA(1), and ARMA(1,1) models, and for nonstationary cases, we use a random walk with i.i.d. errors and a random walk with ARMA(1,1) errors. Our modeling choices are justified by the fact that such low-order processes are commonly used for their parsimony to model stationary and nonstationary economic time series (e.g., Cochrane 1988, Campbell and Mankiw 1987a, 1987b).

To implement interpolation, in each model, we assume that in each $s$-period time segment $t$, $i = 1, 2, ..., s$, only one data point, the benchmark observation, is observed. The remaining $s - 1$ missing observations are generated using linear interpolation. Without loss of generality, we assume that the benchmark observations are $y_{t,s}, t = 1, 2, \ldots$, i.e., the $s^{th}$ observation for each $t$.[6]

## 4. Effect of Interpolation on Shock Persistence of Stationary Series
### 4.1. Stationary AR(1) Process

Consider a stationary AR(1) model, $Y_T = \alpha Y_{T-1} + E_T, T = 1, 2, \ldots$, where $|\alpha| < 1$ and $E_T \sim \text{i.i.d.}(0, \sigma_\varepsilon^2)$. Using the notation above, the process can be rewritten as

$y_{t,i} = \alpha y_{t,i-1} + \varepsilon_{t,i}, t = 1, 2, \ldots,$ and $i = 1, 2, \ldots, s,$ where $\varepsilon_{t,i} \sim \text{i.i.d.}(0, \sigma_\varepsilon^2)$ corresponds to $E_T$ in the same way as $y_{t,i}$ corresponds to $Y_T$.[7]

Setting $i = s$, we have $y_{t,s} = \alpha y_{t,s-1} + \varepsilon_{t,s}$. After successive backward substitution ($s$ times) we obtain

(4) $\quad y_{t,s} = \alpha^s y_{t-1,s} + \sum_{j=1}^{s} \alpha^{s-j} \varepsilon_{t,j}.$

#### 4.1.1. Shock Persistence of the Original, Non-Interpolated Stationary AR(1) Series

To obtain the shock persistence measure $V_y$ for the process in (4), we need to derive the short and the long variances, $\sigma_{1,y}^2$ and $\sigma_{k,y}^2$, respectively. Consider the process $y_{t,s} = \alpha y_{t,s-1} + \varepsilon_{t,s}$ and subtract $y_{t,s-1}$ from both sides. The short variance can be expressed as

(5) $\quad \sigma_{1,y}^2 = \text{var}(y_{t,s} - y_{t,s-1}) = (\alpha - 1)^2 \text{var}(y_{t,s-1}) + \sigma_\varepsilon^2 = \frac{2\sigma_\varepsilon^2}{1+\alpha},$

using the fact that the $\text{var}(y_{t,s-1}) = \frac{\sigma_\varepsilon^2}{1-\alpha^2}$.

To obtain the long variance $\sigma_{k,y}^2$, we rewrite $y_{t,s} - y_{t-1,s}$ by substituting for $y_{t,s}$ from (4). Accordingly, $y_{t,s} - y_{t-1,s} = (\alpha^s - 1)y_{t-1,s} + \sum_{j=1}^{s} \alpha^{s-j} \varepsilon_{t,j}$. The long variance can then be derived as

$\sigma_{k,y}^2 = \text{var}(y_{t,s} - y_{t-1,s}) = \text{var}\left[(\alpha^s - 1)y_{t-1,s} + \sum_{j=1}^{s} \alpha^{s-j} \varepsilon_{t,j}\right]$

---

[6] The formulation is general and it applies to all observations, $y_{t,i}$, $t = 1, 2, \ldots,$ and $i = 1, 2, \ldots, s$. If $s = 1$, then there is no interpolation and thus the original and the interpolated series coincide.

[7] Note that lag of $y_{t,1} = y_{t,0} = y_{t-1,s}$.

(6) $$= (\alpha^s - 1)^2 \text{var}[y_{t-1,s}] + \text{var}\left(\sum_{j=1}^{s} \alpha^{s-j} \varepsilon_{t,j}\right) = \frac{(\alpha^s - 1)^2}{1 - \alpha^2} \sigma_\varepsilon^2 + \frac{1 - \alpha^{2s}}{1 - \alpha^2} \sigma_\varepsilon^2 = \frac{2(1 - \alpha^s)}{1 - \alpha^2} \sigma_\varepsilon^2$$

Note that the above derivation uses the fact that $\text{var}\left(\sum_{j=1}^{s} \alpha^{s-j} \varepsilon_{t,j}\right) = \sigma_\varepsilon^2 \left(\sum_{j=1}^{s} \alpha^{2(s-j)}\right) = \sigma_\varepsilon^2 \left(\sum_{j=1}^{s} \alpha^{2(j-1)}\right)$ and

$\sum_{j=1}^{s} \alpha^{2(j-1)} = \frac{1 - \alpha^{2s}}{1 - \alpha^2}$. The latter equality is used in the proofs below.

The variance ratio for our original, non-interpolated time series $y_{t,s}$ is obtained by substituting (5) and (6) into (3), which yields

(7) $$V_y = \frac{\sigma_{k,y}^2}{k \sigma_{1,y}^2} = \frac{1 - \alpha^s}{s(1 - \alpha)} < 1,$$

where the last inequality, which is in line with Cochrane (1988) and Cogley (1988), is formally proved in Appendix A.I.

*4.1.2. Shock Persistence of the Interpolated Stationary AR(1) Series*

The linearly interpolated series, as denoted by (2) is given by

$x_{t,i} = \frac{i}{s}(y_{t,s} - y_{t-1,s}) + y_{t-1,s} = \frac{i}{s} y_{t,s} + \frac{s-i}{s} y_{t-1,s}$, where $s \geq 2$. With $x_{t,i-1} = \frac{i-1}{s} y_{t,s} + \frac{s-i+1}{s} y_{t-1,s}$ as

a one-period displacement, the short variance of the interpolated series will be the variance of

$x_{t,i} - x_{t,i-1}$ or the variance of $\frac{1}{s}(y_{t,s} - y_{t-1,s})$ given by

(8) $$\sigma_{1,x}^2 = \text{var}(x_{t,i} - x_{t,i-1}) = \left(\frac{1}{s}\right)^2 \text{var}(y_{t,s} - y_{t-1,s}) = \left(\frac{1}{s}\right)^2 \sigma_{k,y}^2 = \frac{2(1 - \alpha^s)}{s^2(1 - \alpha^2)} \sigma_\varepsilon^2.$$

The long variance is the variance of the $s$-period differenced series $x_{t,i} - x_{t-1,i}$.[8] The interpolated series' $k$-period difference can be expressed as follows (recalling that $k = s \geq 2$):

$$x_{t,i} - x_{t-1,i} = \frac{i}{s} y_{t,s} + \frac{s-i}{s} y_{t-1,s} - \frac{i}{s} y_{t-1,s} - \frac{s-i}{s} y_{t-2,s} = i\left(\frac{y_{t,s} - y_{t-1,s}}{s}\right) + (s-i)\left(\frac{y_{t-1,s} - y_{t-2,s}}{s}\right),$$

where the bracketed terms are obtained from equation (4). The long variance is, therefore, given by

$$\sigma_{k,x}^2 = \text{var}\left(x_{t,i} - x_{t-1,i}\right) = i^2 \text{var}\left[\frac{y_{t,s} - y_{t-1,s}}{s}\right] + (s-i)^2 \text{var}\left[\frac{y_{t-1,s} - y_{t-2,s}}{s}\right]$$

$$+ 2i(s-i)\text{cov}\left[\left(\frac{y_{t,s} - y_{t-1,s}}{s}\right), \left(\frac{y_{t-1,s} - y_{t-2,s}}{s}\right)\right].$$

---

[8] We envision the case where annual data are interpolated using decennial "benchmark" observations, consistent with many historical applications of linear interpolation, as discussed earlier. That would fit the interpretation of Cochrane's measure of shock persistence as is often employed by practitioners—a ratio of the variance of the series' cumulative growth over, say, $k = 10$ years, to the variance of the series' 1-year growth.

This shows that the *k*-period growth variance of the interpolated series depends on *i*, implying that it is characterized by *periodic variation*. This is a recurring property in the results we report for the interpolated models we examine. This finding may be related to Romer's (1986a, 1986b, 1986c, 1986d, 1989) argument that interpolation methods used to construct prewar series did, indeed, exaggerate the cyclical patterns in the data.[9]

To eliminate the dependence of the variance on *i*, we compute the expected value of the variance, conditional on *i*. As shown in Appendix A-II, this yields

(9) $\quad \sigma_{k,x}^2 = \left[\dfrac{\sigma_\varepsilon^2}{3s^2(1-\alpha^2)}\right]\left[(1-\alpha^s)(3s^2 + \alpha^s s^2 - \alpha^s + 3)\right].$

The variance ratio statistic for the interpolated AR(1) process is obtained using (3), (8), and (9):

(10) $\quad V_x = \dfrac{\sigma_{k,x}^2}{k\sigma_{1,x}^2} = \dfrac{\left[\dfrac{\sigma_\varepsilon^2}{3s^2(1-\alpha^2)}\right]\left[(1-\alpha^s)(3s^2 + \alpha^s s^2 - \alpha^s + 3)\right]}{s\left[\dfrac{2(1-\alpha^s)}{s^2(1-\alpha^2)}\sigma_\varepsilon^2\right]} = \dfrac{s^2(3+\alpha^s)+(3-\alpha^s)}{6s} > 1.$

We show in Appendix A-III that for the AR(1) process, the variance ratio for the interpolated series is larger than 1.

Two other observations are noteworthy. First, recall that for the original AR(1) series, $V_y < 1$. After linear interpolation, however, we have $V_x > 1$. For a simple random walk process, $V = 1$ (e.g., Cochrane 1988, Cogley 1988). It follows, therefore, that linear interpolation of a stationary AR(1) time series yields a time series that is more persistent than the original series, $V_x > V_y$. Moreover, interpolation makes a series more persistent than a random walk.

Second, the $\lim\limits_{s\to\infty} V_x = \lim\limits_{s\to\infty}\left[\dfrac{s^2(3+\alpha^s)}{6s}\right] + \lim\limits_{s\to\infty}\left[\dfrac{(3-\alpha^s)}{6s}\right]$ diverges, so for a stationary AR(1) process, as the number of sub-periods in between the benchmark observations increases, the interpolated series becomes more persistent. In other words, the persistence of the series obtained through a linear interpolation of a stationary AR(1) series increases with *s*.

---

[9] Our finding here is also consistent with the finding that linearly interpolated series exhibit periodic variation in their moments irrespective of the true nature of the original, non-interpolated series. Dezhbakhsh and Levy (1994) study the effect of linear interpolation on the moments a *trend stationary* series and report that linear interpolation introduces non-stationarity in the series variance-covariance structure. Specifically, they find that the interpolated series' variance, covariance, and the autocorrelation function, all vary with *i* leading to "periodic variation."

## 4.2. Stationary ARMA(1,1) Process

Consider a stationary ARMA(1,1) process $Y_T = \alpha Y_{T-1} + U_T$ where $U_T = E_T + \theta E_{T-1}$ and $\varepsilon_T \sim \text{i.i.d.}(0, \sigma_\varepsilon^2)$, $T = 1, 2, \ldots$, $|\alpha| < 1$ for stationarity, and $|\theta| < 1$ for invertibility. The process can be rewritten as $y_{t,i} = \alpha y_{t,i-1} + u_{t,i}$ and $u_{t,i} = \varepsilon_{t,i} + \theta \varepsilon_{t,i-1}$, where $t = 1, 2, \ldots, i = 1, 2, \ldots, s$, and $\varepsilon_{t,i} \sim \text{i.i.d.}(0, \sigma_\varepsilon^2)$, where $u_{t,i}$ and $\varepsilon_{t,i}$ correspond to $U_T$ and $E_T$ the same way as $y_{t,i}$ corresponds to $Y_T$.

Setting $i = s$, we have $y_{t,s} = \alpha y_{t,s-1} + \varepsilon_{t,s} + \theta \varepsilon_{t,s-1}$, which, after successive backward substitutions $s$ times, yields

$$(11) \quad y_{t,s} = \alpha^s y_{t-1,s} + \sum_{j=0}^{s-2} \alpha^j \left( \varepsilon_{t,s-j} + \theta \varepsilon_{t,s-j-1} \right) + \alpha^{s-1} \left( \varepsilon_{t,1} + \theta \varepsilon_{t-1,s} \right),$$

where the last term is written separately because its subscripts do not correspond to the summation index.

### 4.2.1. Shock Persistence of the Original, Non-Interpolated Stationary ARMA(1,1) Series

Because $y_{t,s} - y_{t,s-1} = (\alpha - 1) y_{t,s-1} + \varepsilon_{t,s} + \theta \varepsilon_{t,s-1}$, we have

$$(12) \quad \begin{aligned} \sigma_{1,y}^2 &= \text{var}(y_{t,s} - y_{t,s-1}) = \text{var}\left[(\alpha-1) y_{t,s-1} + \varepsilon_{t,s} + \theta \varepsilon_{t,s-1}\right] \\ &= (\alpha-1)^2 \sigma_y^2 + (1+\theta^2) \sigma_\varepsilon^2 + 2(\alpha-1) \theta \sigma_\varepsilon^2, \end{aligned}$$

where the last term in equation (12) is the covariance of $y_{t,s-1}$ and $\varepsilon_{t,s-1}$, and $\sigma_y^2 = \dfrac{(1+\theta^2+2\alpha\theta)}{1-\alpha^2} \sigma_\varepsilon^2$ is the variance of the process $y_t$.

The long variance can be obtained by subtracting $y_{t-1,s}$ from both sides of (11), which yields $y_{t,s} - y_{t-1,s} = (\alpha^s - 1) y_{t-1,s} + \sum_{j=0}^{s-2} \alpha^j \left( \varepsilon_{t,s-j} + \theta \varepsilon_{t,s-j-1} \right) + \alpha^{s-1} \left( \varepsilon_{t,1} + \theta \varepsilon_{t-1,s} \right)$. Then, for $k = s \geq 2$, we have

$$(13) \quad \begin{aligned} \sigma_{k,y}^2 &= \text{var}(y_{t,s} - y_{t-1,s}) \\ &= (\alpha^s - 1)^2 \sigma_y^2 + (1+\theta^2) \sigma_\varepsilon^2 \left( \sum_{j=1}^{s} \alpha^{2(j-1)} \right) + 2\theta \sigma_\varepsilon^2 \left( \sum_{j=1}^{s-1} \alpha^j \alpha^{j-1} \right) + 2(\alpha^s - 1) \alpha^{s-1} \theta \sigma_\varepsilon^2 \\ &= (\alpha^s - 1)^2 \sigma_y^2 + (1+\theta^2 + 2\alpha\theta) \sigma_\varepsilon^2 \left( \frac{1-\alpha^{2s}}{1-\alpha^2} \right) - 2\alpha^{s-1} \theta \sigma_\varepsilon^2 \end{aligned}$$

The variance ratio is then computed using (12) and (13):

$$V_y = \frac{\sigma_{k,y}^2}{k\sigma_{1,y}^2} = \left(\frac{1}{s}\right) \left[ \frac{(\alpha^s-1)^2 \sigma_y^2 + (1+\theta^2)\sigma_\varepsilon^2 \left(\frac{1-\alpha^{2s}}{1-\alpha^2}\right) + 2\theta\sigma_\varepsilon^2 \left(\frac{\alpha-\alpha^{2s+1}}{1-\alpha^2}\right) - 2\alpha^{s-1}\theta\sigma_\varepsilon^2}{(\alpha-1)^2 \sigma_y^2 + (1+\theta^2)\sigma_\varepsilon^2 + 2(\alpha-1)\theta\sigma_\varepsilon^2} \right]$$

(14)

$$= \left(\frac{1}{s}\right) \left[ \frac{(1+\theta^2+2\alpha\theta) - \alpha^{s-1}(1+\alpha\theta)(\alpha+\theta)}{(1+\theta^2+2\alpha\theta) - (1+\alpha\theta)(\alpha+\theta)} \right]$$

The derivation details for the long variance and variance ratio are provided in Appendix A-IV.

This expression does not lend itself to easily assessing the value of the variance ratio. We can show, however, that if the first-order autocorrelation $\rho_1 < 0$, then $V_y < 1$. If $\rho_1 > 0$, then $V_y < 1$ as long as $\rho_1 < \frac{s-1}{s-\alpha^{s-1}}$ (see Appendix A-V for the proof). So, the variance ratio is less than one when the first-order autocorrelation of the ARMA(1,1) series is either negative or positive but less than $\frac{s-1}{s-\alpha^{s-1}}$. But over what segment of the parameter space $(\alpha \times \theta)$ is the latter condition satisfied?

The limiting value of $\frac{s-1}{s-\alpha^{s-1}}$ is 1 when $s$ is large. In such cases, the variance ratio will always be less than 1. For example, in the case of decennial data with $s = 10$, we have $0.81 < \frac{s-1}{s-\alpha^{s-1}} < 1.00 \ \forall \alpha \in (-1,1)$. Therefore, as long as $\rho_1 < 0.81$, the variance ratio is less than 1.

To examine the segment of the parameter space over which $\rho_1$ is positive but larger than $\frac{s-1}{s-\alpha^{s-1}}$ (which yields a variance ratio larger than 1), we compute the variance ratio for different values of $|\alpha|<1$, $|\theta|<1$, and $s \geq 2$. Note that because $\alpha \in (-1,1)$, $\theta \in (-1,1)$ and $s \geq 2$, the denominator of the ratio is $s\left[(1+\theta^2+2\alpha\theta) - (1+\alpha\theta)(\alpha+\theta)\right] \neq 0$. Therefore, the variance ratio function $V_y(\alpha,\theta)$ is continuous in its domain.

Because the variance ratio here depends on $\alpha, \theta,$ and $s$ in a complicated manner, we plot the ratio over the relevant segment of the parameter space for various values of $s$. Figure 1 shows a 3D surface plot of $V_y(\alpha,\theta)$ for the non-interpolated ARMA(1,1) series for $\alpha \in (-1,1)$ and $\theta \in (-1,1)$ for various values of $s \geq 2$.[10] According to the plots, the range of the function is positive, as it should be because the numerator and the denominator of the function are variances. However, the function's upper bound, as indicated by the highest value along the vertical axis, changes with $s$ but more importantly, over the domain defined by $\alpha \in (-1,1)$ and $\theta \in (-1,1)$.[11]

---

[10] Incremental information from figures for other values of $s$ adds very little to the findings and does not alter our conclusions.

The function's upper bounds, as measured by the vertical axis—the largest value that the variance ratio can achieve over the relevant region of the parameter space—are approximately 1.5 for $s = 2$, and 1.8 for $s = 4$. Nonetheless, the surface achieves a height of less than 1 for the vast part of the domain $\alpha \in (-1,1)$ and $\theta \in (-1,1)$. A surface lift is noticed in the northeast corner that corresponds to the boundary values where $\alpha$ and $\theta$ approach 1. Variance ratio values larger than one are only observed in this corner. The surface has a similar pattern across all values of $s$ except for the location of this lift, which varies slightly with $s$. The lift is observed over smaller and smaller regions of the parameter space as $s$ increases. The change is more noticeable when comparing $s = 2$ to $s = 25$.[12] This suggests that in most cases, the variance ratio for a non-interpolated ARMA(1,1) series is less than one.

*4.2.2. Shock Persistence of the Interpolated Stationary ARMA(1,1) Series*

The interpolated series $x_{t,i}$ is given by $x_{t,i} = \frac{i}{s}(y_{t,s} - y_{t-1,s}) + y_{t-1,s}$, with a one-period lag displacement $x_{t,i-1} = \frac{i-1}{s} y_{t,s} + \frac{s-i+1}{s} y_{t-1,s}$, and thus $x_{t,i} - x_{t,i-1} = \frac{1}{s}(y_{t,s} - y_{t-1,s})$. Substituting for $y_{t,s}$ from equation (11), we obtain the short variance

(15)
$$\sigma_{1,x}^2 = \text{var}(x_{t,i} - x_{t,i-1}) = \left(\frac{1}{s}\right)^2 \text{var}\left[(\alpha^s - 1) y_{t-1,s} + \sum_{j=0}^{s-2} \alpha^j (\varepsilon_{t,s-j} + \theta \varepsilon_{t,s-j-1}) + \alpha^{s-1}(\varepsilon_{t,1} + \theta \varepsilon_{t-1,s})\right]$$
$$= \frac{2\sigma_\varepsilon^2}{s^2(1-\alpha^2)}\left[(1-\alpha^s)(1+\alpha\theta+\theta^2) + \alpha\theta(1-\alpha^{s-2})\right]$$

Proofs are presented in Appendix A-VI.

The $s$-period difference is given by $i\left(\frac{y_{t,s} - y_{t-1,s}}{s}\right) + (s-i)\left(\frac{y_{t-1,s} - y_{t-2,s}}{s}\right)$, as shown in section 4.1.2. Therefore, the long variance equals

---

[11] We use "upper bound" and not "maximum" because the function $V_y(\alpha, \theta)$ has no maximum. It is a continuous function that is defined on the rectangular domain, a subset of $\mathbb{R}^2$, formed by $\alpha \in (-1,1)$ and $\theta \in (-1,1)$, which is an open set.

[12] Notes: (a) The vertical axis scale used in Figure 1, as well as in Figures 2, 3, and 4, was set by the Mathematica software. (b) The values along the vertical axis at the origin are not always 0. (c) The numbers along the vertical axis correspond to the long tick marks. (d) We have produced the plots for the following values of $s$: 2, 3, 4, 5, 6, 7, 8, 9, 10, 15, 20, 25, and 30. For brevity, however, we only present the plots for s = 2, 4, 5, 10, 15, and 25.

$$\sigma_{k,x}^2 = \text{var}(x_{t,i} - x_{t-1,i})$$

$$= i^2 \text{var}\left[\frac{y_{t,s} - y_{t-1,s}}{s}\right] + (s-i)^2 \text{var}\left[\frac{y_{t-1,s} - y_{t-2,s}}{s}\right] + 2i(s-i)\text{cov}\left[\left(\frac{y_{t,s} - y_{t-1,s}}{s}\right), \left(\frac{y_{t-1,s} - y_{t-2,s}}{s}\right)\right]$$

$$= \left[s^2 + 2i^2 - 2si\right]\left[\frac{2\sigma_\varepsilon^2}{s^2(1-\alpha^2)}\right]\left[(1-\alpha^s)(1+\alpha\theta+\theta^2) + \alpha\theta(1-\alpha^{s-2})\right]$$

$$+ \frac{2i(s-i)}{s^2}\left[(2\alpha^{s-1} - \alpha^{2s-1})\gamma_1 - \gamma_0\right]$$

where the autocovariances in the last equation are given by $\gamma_0 = \dfrac{1+2\alpha\theta+\theta^2}{1-\alpha^2}\sigma_\varepsilon^2$, and

$\gamma_1 = \dfrac{(1+\alpha\theta)(\alpha+\theta)}{1-\alpha^2}\sigma_\varepsilon^2$. See Appendix A-VI for derivation details.

The expected value of the long variance is given by

(16)
$$\sigma_{k,x}^2 = \left(\frac{2s^2+1}{3}\right)\left[\frac{2\sigma_\varepsilon^2}{s^2(1-\alpha^2)}\right]\left[(1-\alpha^s)(1+\alpha\theta+\theta^2) + \alpha\theta(1-\alpha^{s-2})\right]$$

$$+ \left(\frac{s^2-1}{3s^2}\right)\left[(2\alpha^{s-1} - \alpha^{2s-1})\frac{(1+\alpha\theta)(\alpha+\theta)}{1-\alpha^2}\sigma_\varepsilon^2 - \left(\frac{1+2\alpha\theta+\theta^2}{1-\alpha^2}\right)\sigma_\varepsilon^2\right]$$

See Appendix A-VI for the above (and the following) derivation details.

The ratio of (15) and (16) divided by $s$ yields the variance ratio.

(17) $\quad V_x = \dfrac{\sigma_{k,x}^2}{k\sigma_{1,x}^2} = \left(\dfrac{2s^2+1}{3s}\right) + \left(\dfrac{s^2-1}{6s}\right)\left[\dfrac{\alpha^{s-1}(2-\alpha^s)(1+\alpha\theta)(\alpha+\theta) - (1+2\alpha\theta+\theta^2)}{(1-\alpha^s)(1+2\alpha\theta+\theta^2) + \alpha\theta(1-\alpha^{s-2})}\right]$

If we set $\theta = 0$, then (17) simplifies to $V_x = \dfrac{s^2(3+\alpha^s)+(3-\alpha^s)}{6s}$ which is the variance ratio of an interpolated AR(1) process, given in equation (10), in section 4.1.2.

Because (17) depends on the parameters of the model as well as $s$ in a complicated manner, we plot it over the relevant segment of the parameter space for various values of $s$. Figure 2 shows a 3D surface plot of $V_y(\alpha,\theta)$ for various values of $s \geq 2$. According to the plots, the range of the function is positive, but the function's upper bound, as indicated by the highest value along the vertical axis, changes with $s$ as well as over the domain $\alpha \in (-1,1)$ and $\theta \in (-1,1)$.

Unlike Figure 1, here, variance ratio values larger than 1 are prevalent over much of the parameter space. These values increase significantly with $s$, because a larger $s$ more interpolated observations and thus more contamination. For example, the largest value that the variance ratio attains is around 1.5 for $s = 2$ but it equals 16 for $s = 25$. This is due to the order of $s$ in (17). We also notice that the contour of the surface is different for odd and even values of $s$. Also, for odd values of $s$, the ratio is more affected by a change in the AR parameter $\alpha$ than a change in the MA parameter $\theta$.

Finally, to assess the impact of interpolation on the variance ratio of the ARMA(1,1) model, we compute the variance ratio for the original series and its interpolated version, for various values of $\alpha, \theta,$ and $s$. The parameter values chosen represent the relevant region of the parameter space that satisfy stationarity and convertibility conditions, and two values of $s$, 4 and 10, which are a reasonable representation of historical uses of interpolation (quarterly and decennial).[13]

The results of our numerical evaluation of the variance ratios for the two series ($V_x$ and $V_y$) are presented in Table 1 (for $s=4$) and Table 2 (for $s=10$). Results reported in the tables indicate that interpolation increases the value of the variance ratio. For example, for all parameter values examined, the ratio for the non-interpolated ARMA(1,1) series is smaller than the ratio for the interpolated ARMA(1,1) series. The results are more pronounced when $s$ increases from 4 to 10. Note that the variance ratio for the original series, $V_y$, has some values that are larger than one for smaller $s$ values. This is not surprising given that the variance ratio limiting values are approximations achieved when $s$ approaches infinity (See, e.g., Cochran, 1988, pages 895, 897, 906, and 908, and his discussion about parameter $k$, which we denote by $s$). Tables 1 and 2 show that, in general, the $V_y$ values are smaller for larger $s$, as expected.

Moreover, in all interpolated ARMA(1,1) cases, the values of the variance ratio are larger than 1, and often significantly so. Such large variance ratio values are usually expected from series with a random walk and not stationary ARMA series (Cochrane, 1988). For the non-interpolated ARMA(1,1) series, however, the variance ratio values are less than 1 over most relevant parameter values, and they decrease as $s$ increases from 4 to 10. These results are consistent with, and further strengthen, our theoretical finding for AR(1) series in previous sections that interpolation of a series increases its shock persistence.

### 4.3. MA(1) Process

Using the results in section 4.2, we can set the AR parameter $\alpha$ equal to 0 to examine the impact of linear interpolation on series generated by a stationary MA(1) process.

### 4.3.1. Shock Persistence of the Original, Non-Interpolated MA(1) Series

Consider an MA(1) process $y_{t,i} = \varepsilon_{t,i} + \theta \varepsilon_{t,i-1}$, where $t=1,2,..., i=1,2,...,s$, and $\varepsilon_{t,i} \sim \text{i.i.d.}(0, \sigma_\varepsilon^2)$. By substituting the autoregressive parameter $\alpha = 0$ in equations (12) and (13), we obtain the short and long variances for the original, non-interpolated MA(1) process

$$\sigma_{1,y}^2 = \left[2(1+\theta^2) - 2\theta\right]\sigma_\varepsilon^2$$

and

$$\sigma_{k,y}^2 = \left[2(1+\theta^2)\right]\sigma_\varepsilon^2$$

---

[13] Note that the long difference values commonly used are 10, 20, and 25, but we observed that the variance ratios for the non-interpolated and interpolated series do not change much for values of $s$ above 10.

The variance ratio statistic of the original, non-interpolated MA(1) process is, therefore, given by

$$V_y = \frac{\sigma_{k,y}^2}{k\sigma_{1,y}^2} = \left(\frac{1}{s}\right)\left(\frac{1+\theta^2}{1+\theta^2 - \theta}\right) < 1$$

where the inequality holds because $s \geq 2$ and $|\theta| < 1$. The result that $V_y < 1$, is consistent with the fact that the underlying MA(1) process is stationary.

*4.3.2. Shock Persistence of the Interpolated MA(1) Series*

To derive the variance ratio for the linearly interpolated above MA(1) series, we set the autoregressive parameter $\alpha = 0$ in equations (15) and (16), yielding the short and long variances of the interpolated MA(1) series

$$\sigma_{1,x}^2 = \frac{2}{s^2}\left[(1+\theta^2)\right]\sigma_\varepsilon^2$$

and

$$\sigma_{k,x}^2 = \left(\frac{s^2+1}{s^2}\right)(1+\theta^2)\sigma_\varepsilon^2$$

The variance ratio statistic of the interpolated MA(1) series is, therefore, given by

$$V_x = \frac{\sigma_{k,x}^2}{k\sigma_{1,x}^2} = \frac{s^2+1}{2s} > 1 \text{ for } s \geq 2.$$

It follows that linear interpolation turns the stationary MA(1) process into a nonstationary process that is more persistent than a random walk. Moreover, $\lim_{s\to\infty} V_x = \lim_{s\to\infty}\left(\frac{s^2+1}{2s}\right)$ diverges. Thus, as the number of interpolated sub-periods between the benchmark observations increases, the more persistent the resulting interpolated series will become.

**5. Effect of Interpolation on Shock Persistence of Nonstationary Series**

Next, we examine the effect of interpolation on the shock persistence of a nonstationary (difference stationary) series with i.i.d. errors and a nonstationary series with ARMA(1,1) errors. The advantage of these models is that they are widely used in modeling macroeconomic time series, and also, as Cochrane (1988) argues, they can be shown to be equivalent to mixed models with a random walk and a stationary component.

### 5.1. Nonstationary Series with i.i.d. Errors

Consider $Y_T = \mu + Y_{T-1} + E_T$, $T = 1, 2, ...$ where $\mu$ is the drift, and $E_T$ is the error term. We rewrite it as $y_{t,i} = \mu + y_{t-1,i} + \varepsilon_{t,i}$, $i = 1, 2, ..., s$, $s \geq 2$, where $\varepsilon_{t,i} \sim \text{i.i.d.}(0, \sigma_\varepsilon^2)$ corresponds to $E_T$ the same way as $y_{t,i}$ corresponds to $Y_T$. After successive substitutions, $y_{t,i}$ can be written as

$$(18) \quad y_{t,i} = s\mu + y_{t-1,i} + \sum_{j=1}^{s} \varepsilon_{t,j}, \quad i = 1, 2, ..., s.$$

#### 5.1.1. Shock Persistence of the Original, Non-Interpolated Nonstationary Series with i.i.d. Errors

The short variance is

$$\sigma_{1,y}^2 = \text{var}(y_{t,i} - y_{t,i-1}) = \text{var}(\varepsilon_{t,i}) = \sigma_\varepsilon^2,$$

while the long variance is given by

$$\sigma_{k,y}^2 = \text{var}(y_{t,i} - y_{t-1,i}) = \text{var}\left(\sum_{j=1}^{s} \varepsilon_{t,j}\right) = s\sigma_\varepsilon^2.$$

The variance ratio for the non-interpolated random walk is, therefore,

$$V_y = \frac{\sigma_{k,y}^2}{k\sigma_{1,y}^2} = \frac{s\sigma_\varepsilon^2}{k\sigma_\varepsilon^2} = 1,$$

which is in line with the results reported by Cochrane (1988), Cogley (1990), Levy and Dezhbakhsh (2003), and the studies cited therein.

#### 5.1.2. Shock Persistence of an Interpolated Nonstationary Series with i.i.d. Errors

Linearly interpolated series are given by $x_{t,i} = \frac{i}{s} y_{t,s} + \frac{s-i}{s} y_{t-1,s}$, and its one-period lag equals $x_{t,i-1} = \frac{i-1}{s} y_{t,s} + \frac{s-i+1}{s} y_{t-1,s}$. Thus, $x_{t,i} - x_{t,i-1} = \frac{1}{s}(y_{t,s} - y_{t-1,s})$. Then, (18) yields the short variance of the interpolated series:

$$\sigma_{1,x}^2 = \text{Var}(x_{t,i} - x_{t,i-1}) = \left(\frac{1}{s}\right)^2 \text{Var}(y_{t,s} - y_{t-1,s}) = \left(\frac{1}{s}\right)^2 s\sigma_\varepsilon^2 = \frac{1}{s}\sigma_\varepsilon^2.$$

The long variance is obtained by expressing $x$ in terms of $y$ and using (18):

$$\sigma_{k,x}^2 = \text{Var}(x_{t,i} - x_{t-1,i}) = \text{Var}\left[\frac{i}{s}(y_{t,s} - y_{t-1,s}) + \frac{s-i}{s}(y_{t-1,s} - y_{t-2,s})\right]$$

$$= \text{Var}\left[\frac{i}{s}\left(s\mu + \sum_{j=1}^{s}\varepsilon_{t,j}\right) + \frac{s-i}{s}\left(s\mu + \sum_{j=1}^{s}\varepsilon_{t-1,j}\right)\right] = \left(\frac{i}{s}\right)^2 s\sigma_\varepsilon^2 + \left(\frac{s-i}{s}\right)^2 s\sigma_\varepsilon^2$$

$$= \frac{i^2 + (s-i)^2}{s}\sigma_\varepsilon^2.$$

The expected value of the variance, conditional on $i$, is given by

$$\sigma_{k,x}^2 = \frac{\sum_{i=1}^{s}\left[\frac{i^2+(s-i)^2}{s}\right]\sigma_\varepsilon^2}{s} = \frac{\sigma_\varepsilon^2}{s^2}\sum_{i=1}^{s}\left[i^2+(s-i)^2\right] = \frac{\sigma_\varepsilon^2}{s^2}\left(s^3 + 2\frac{s(s+1)(2s+1)}{6} - 2s\frac{s(s+1)}{2}\right)$$

$$= \left(\frac{2s^2+1}{3s}\right)\sigma_\varepsilon^2$$

Therefore, the variance ratio of the interpolated random walk is given by

(19) $\quad V_x = \dfrac{\sigma_{k,x}^2}{k\sigma_{1,x}^2} = \dfrac{2s^2+1}{3s}.$

Thus, for $s \geq 2$, $V_x > V_y = 1$.[14] Moreover, the higher the value of $s$, the greater is $V_x$. In other words, the persistence of a series obtained through linear interpolation increases with the extent of interpolation. This implies that by linearly interpolating a difference-stationarity series, we generate a new series which are characterized by greater shock persistence than the original series.

### 5.2. Nonstationary Series with ARMA(1,1) Errors

Next, we generalize the random walk model by allowing the error terms to follow an ARMA(1,1) process: $Y_T = \mu + Y_{T-1} + U_T$, $T = 1,2,...$ where $\mu$ is the drift parameter and $U_T$ follows an ARMA(1,1) structure. We rewrite the model as $y_{t,i} = \mu + y_{t-1,i} + u_{t,i}$, where $y_{t,i}$ and $u_{t,i}$ correspond to $Y_T$ and $U_T$, and the error term follows ARMA(1,1), $u_{t,i} = \alpha u_{t,i-1} + \varepsilon_{t,i} + \theta\varepsilon_{t,i-1}$ where $\varepsilon_{t,i} \sim \text{i.i.d.}(0,\sigma_\varepsilon^2)$, and the stationarity and the invertibility conditions, $|\theta|<1$ and $|\alpha|<1$, are satisfied. After successive substitution, we can write

---

[14] Substituting $s=1$ into (19), which is equivalent to not interpolating the series, yields $V_x = V_y = 1$.

(20) $\quad y_{t,s} = s\mu + y_{t-1,s} + \sum_{j=1}^{s} u_{t,j}$.

*5.2.1. Shock Persistence of a Non-Interpolated Nonstationary Series with ARMA(1,1) Errors*

We obtain the one-period and the *s*-period variances for this series as follows: The short variance is

$$\sigma_{1,y}^2 = \text{var}(y_{t,i} - y_{t,i-1}) = \text{var}(u_{t,i}) = \frac{1+\theta^2+2\alpha\theta}{1-\alpha^2}\sigma_\varepsilon^2 = \gamma_0,$$

where $\gamma_0$ is the variance of the ARMA(1,1) error process.

The long variance attains a more complicated form.

$$\sigma_{k,y}^2 = \text{var}(y_{t,i} - y_{t-1,i}) = \text{var}\left(\sum_{j=1}^{s} u_{t,j}\right) = s\gamma_0 + 2\sum_{j=1}^{s-1}(s-j)\gamma_j,$$

where $\gamma_1 = \frac{(1+\alpha\theta)(\alpha+\theta)}{1-\alpha^2}\sigma_\varepsilon^2$ and $\gamma_j = \alpha\gamma_{j-1}$, $j \geq 2$, are the covariance functions for the error term. The long variance expression is equivalent to equation (9) in Cochrane (1988, p. 906), and to equation (1) in Cogley (1990, p. 503), for a difference-stationary process.

As we show in Appendix B-I, the variance ratio statistic can be expressed as

$$(21) \quad V_y = \frac{\sigma_{k,y}^2}{k\sigma_{1,y}^2} = \frac{s\gamma_0 + 2\gamma_1\sum_{j=1}^{s-1}(s-j)\alpha^{j-1}}{s\gamma_0} = 1 + \frac{2\rho_1}{s}\left[\sum_{j=1}^{s-1}(s-j)\alpha^{j-1}\right] = 1 + \frac{2\rho_1}{s}\left[\frac{s(1-\alpha)-(1-\alpha^s)}{(1-\alpha)^2}\right]$$

where $\rho_1$, the first-order autocorrelation of the error process $u_{t,i}$, is given by

$$\rho_1 = \frac{\gamma_1}{\gamma_0} = \frac{(1+\alpha\theta)(\alpha+\theta)}{1+\theta^2+2\alpha\theta}.\ [15]$$

In Appendix B-II, we prove the following properties of (21). First, the bracketed expression in (21) is positive because $|\alpha|<1$ and $s>1$, implying that $V_y > 1$ if $\rho_1 > 0$, which is the case for many economic series. Moreover, $V_y < 1$ if $\rho_1 < 0$, and $V_y = 1$ if $\rho_1 = 0$, which is the case of a random walk. Thus, the size of the variance ratio statistic depends on the sign of the first-order autocorrelation: the greater the first-order autocorrelation of the ARMA(1,1) errors $\rho_1$, the more shock persistent the nonstationary time series is.

Second, if $\alpha = -\theta$, (21) implies that $\rho_1 = 0$ and thus $V_y = 1$. Also, If $\alpha = 0$, then the error process is an MA(1) and $\rho_1 = \frac{\theta}{1+\theta^2}$. In this case, the variance ratio is $V_y = 1 + 2\left(\frac{\theta}{1+\theta^2}\right)\left(\frac{s-1}{s}\right)$.

---

[15] Equation (21) is equivalent to a similar equation that Cochrane (1988) derives in his Appendix A, page 917.

But because $s > 1$, we have $\frac{s-1}{s} > 0$, and since $|\theta| < 1$, the variance ratio satisfies $0 < V_y < 2$ and, therefore, the size of $V_y$ depends on the sign and magnitude of $\theta$. In particular, $V_y > 1$ if $\theta > 0$, $V_y = 1$ if $\theta = 0$ (which is the case of a random walk), and $V_y < 1$ if $\theta < 0$.

Third, in a more general case with $s = 2$, $\alpha > 0$, and $\theta > 0$, we have $V_y > 1$, if $\alpha < 0$ and $\theta < 0$, then $V_y < 1$, and if $\alpha = 0$ and $\theta = 0$, then $V_y = 1$.

To supplement the above theoretical assessment, we compute the variance ratio for the original, non-interpolated series over the relevant regions of the parameter space where the stationarity and invertibility conditions hold. Figure 3 presents these as 3D surface plots of the variance ratio function $V_y(\alpha, \theta)$ for various values of $s \geq 2$. According to the plots, the range of the function is positive, but the function's upper bound, as indicated by the highest value along the vertical axis, changes with $s$ as well as over the domain of $\alpha \in (1, -1)$ and $\theta \in (1, -1)$.

Consider first the diagonal that connects the southern corner of the figure (where $\alpha$ is close to $-1.0$ and $\theta$ is close to $1.0$) to the northern corner (where $\alpha$ is close to $1.0$ and $\theta$ is close to $-1.0$). Along this diagonal, the value of the variance ratio equals 1. That is because at these points $\alpha + \theta = 0$ and thus the RHS term in the expression for $V_y$ vanishes. For the parameter value combinations that lie on the northeast side of the diagonal, the variance ratio is greater than 1, while for the parameter value combinations that lie on the southwest side of the diagonal, the variance ratio is less than 1.

Next, consider the effect of an increase in $s$ from 2 to 25. As the plots indicate, the values of the variance ratio in the northeast half of the plots increase with $s$. For example, the variance ratio values' upper bounds as measured along the vertical axis, is around 2.0 for $s = 2$ but it increases to 20 for $s = 25$. A surface lift is noticed along the northeast side of the figures that corresponds to the boundary value where $\alpha$ approaches 1. The surface lift gets sharper and steeper as the value of $s$ increases. The variance ratio of the series is particularly high for values of $\alpha$ near 1.

*5.2.2. Shock Persistence of the Interpolated Nonstationary Series with ARMA(1,1) Errors*

Recall from section 5.1.2 that $x_{t,i} - x_{t,i-1} = \frac{1}{s}(y_{t,s} - y_{t-1,s})$, and its the short variance

$$\sigma_{1,x}^2 = \left(\frac{1}{s}\right)^2 \text{var}(y_{t,s} - y_{t-1,s}) = \left(\frac{1}{s}\right)^2 \sigma_{k,y}^2 = \left(\frac{1}{s}\right)^2 \left[s\gamma_0 + 2\gamma_1 \sum_{j=1}^{s-1}(s-j)\alpha^{j-1}\right].$$

The derivations for all results reported in this section are presented in Appendix B-III.

The long variance can be obtained using the facts that

$$x_{t,i} = \frac{i}{s}\left(s\mu + y_{t-1,s} + \sum_{j=1}^{s} u_{t,j}\right) + \frac{s-i}{s} y_{t-1,s} = i\mu + \left(s\mu + y_{t-2,s} + \sum_{j=1}^{s} u_{t-1,j}\right) + \frac{i}{s}\left(\sum_{j=1}^{s} u_{t,j}\right), \text{ and}$$

$$x_{t-1,i} = i\mu + y_{t-2,s} + \frac{i}{s}\left(\sum_{j=1}^{s} u_{t-1,j}\right). \text{ Therefore, } x_{t,i} - x_{t-1,i} = s\mu + \frac{s-i}{s}\left(\sum_{j=1}^{s} u_{t-1,j}\right) + \frac{i}{s}\left(\sum_{j=1}^{s} u_{t,j}\right), \text{ and thus}$$

$$\sigma_{k,x}^2 = \text{var}\left[s\mu + \frac{s-i}{s}\left(\sum_{j=1}^{s} u_{t-1,j}\right) + \frac{i}{s}\left(\sum_{j=1}^{s} u_{t,j}\right)\right]$$

$$= \left[\frac{(s-i)^2 + i^2}{s}\right]\left[s\gamma_0 + 2\gamma_1 \sum_{j=1}^{s-1}(s-j)\alpha^{j-1}\right] + 2\left(\frac{s-i}{s}\right)\left(\frac{i}{s}\right)\left[\sum_{j=1}^{s-1}(s-j)\gamma_{s-j} + \sum_{j=0}^{s-1}(s-j)\gamma_{s+j}\right].$$

The expected value of the variance, conditional on $i$, is given by

$$\sigma_{k,x}^2 = \left(\frac{2s^2+1}{3s^2}\right)\left[s\gamma_0 + 2\gamma_1 \sum_{j=1}^{s-1}(s-j)\alpha^{j-1}\right] + \left(\frac{s^2-1}{3s^2}\right)\left[\sum_{j=1}^{s-1}(s-j)\gamma_{s-j} + \sum_{j=0}^{s-1}(s-j)\gamma_{s+j}\right].$$

The variance ratio, therefore, is then given by

$$V_x = \frac{\sigma_{k,x}^2}{k\sigma_{1,x}^2} = \left(\frac{2s^2+1}{3s}\right) + \left(\frac{s^2-1}{3s}\right)\rho_1 \left[\frac{\sum_{j=1}^{s-1}(s-j)\alpha^{s-j-1} + \sum_{j=0}^{s-1}(s-j)\alpha^{s+j-1}}{s + 2\rho_1 \sum_{j=1}^{s-1}(s-j)\alpha^{j-1}}\right],$$

where, the equality $\rho_1 = \frac{\gamma_1}{\gamma_0}$ is used to obtain the above expression. The variance ratio can be further simplified, as shown in Appendix B-III. This yields

(22) $$V_x = \left(\frac{2s^2+1}{3s}\right) + \frac{(1+\alpha\theta)(\alpha+\theta)(s^2-1)(1-\alpha^s)^2}{3s^2(1-\alpha)^2(1+\theta^2+2\alpha\theta) + 6s(1+\alpha\theta)(\alpha+\theta)\left[s(1-\alpha)-(1-\alpha^s)\right]}.$$

Note that in the special case with $\alpha = 0$ and $\theta = 0$, we have that $\rho_1 = 0$ and therefore the variance ratio becomes $V_x = \left(\frac{2s^2+1}{3s}\right)$, which is identical to (19), the variance ratio we obtained in the case of an interpolated random walk.

The variance ratio values in (22) depend on the parameters of the ARMA(1,1) model and $s$ in a complicated manner. We, therefore, plot the ratio over the relevant segment of the parameter space for various values of $s$ to gain insight into the variance ratio pattern. Figure 4 presents a 3D surface plots of $V_y(\alpha,\theta)$ for various values of $s \geq 2$. As the plots show, irrespective of the value of $\alpha$ and $\theta$, for all values of $s \geq 2$, the variance ratio statistic attains values that exceed 1, which means that the interpolated nonstationary series with ARMA(1,1) errors, will always exhibit more shock persistence than a pure random walk. The variance ratio statistics' upper bound, as indicated by the highest value along the vertical axis, changes with $s$ as well as over the domain of $\alpha \in (1,-1)$ and $\theta \in (1,-1)$.

The variance ratio attains its highest values along the upper edge, where $\alpha$ is near 1. The contour of the surface is different for odd and even values of $s$. The lowest values that the variance ratio attains vary correspondingly. If $s$ is even, then the lowest value of the variance ratio occurs when $\alpha = 0$. If $s$ is odd, then the lowest value of the variance ratio occurs when $\alpha$ is near $-1$. The surface slope changes along the top and bottom edges, which correspond to the boundary values of $\alpha$, in the proximity of 1 and $-1$. The surface slope changes are sharper as the value of $s$ increases.

To assess the impact of interpolation on the variance ratio of a nonstationary series with ARMA(1,1) errors, we evaluate the variance ratio for the original series and its interpolated version, for various values of $\alpha, \theta,$ and $s$. As before, the parameter values are chosen in an encompassing fashion, satisfying stationarity and convertibility conditions, and two $s$ values of 4 and 10.

Numerical evaluation of the variance ratios for the two series ($V_x$ and $V_y$) are presented in Table 3 (for $s = 4$) and Table 4 (for $s = 10$). Results reported in both tables indicate that interpolation increases the value of the variance ratio for a difference-stationary ARMA(1,1) model. Moreover, the difference between the variance ratios for the non-interpolated and interpolated series becomes more pronounced as $s$ increases from 4 to 10.

Moreover, when the series is a random walk with no ARMA component ($\alpha = \theta = 0$), then the variance ratio for the original series is 1, and for the interpolated series it is significantly larger than 1. For positive $\alpha$ and $\theta$, while the variance ratio for the original series is larger than 1 over the most relevant region of the parameter space, the interpolated series exhibits even more shock persistence as its variance ratio is larger. These results are consistent with, and further strengthen, our theoretical finding that interpolation increases the shock persistence of a series.

**6. Effect of Linear Interpolation on Variance Components and Volatility of Time Series**

In previous sections, we reported that for all five DGPs we study $V_x > V_y$, that is, the variance ratio of the interpolated series $x$ always exceeds the variance ratio of the original, non-interpolated series $y$, irrespective of the DGP, and regardless of whether or not the original series is stationary or nonstationary.

To further understand the reason behind this result, we examine the effect of linear interpolation on the numerator and denominator of the variance ratio of both the original and interpolated series, which are the short and long variances, respectively, $V_y = \dfrac{\sigma_{k,y}^2}{k\sigma_{1,y}^2}$ and $V_x = \dfrac{\sigma_{k,x}^2}{k\sigma_{1,x}^2}$.

In this section we show that in all cases, $\sigma_{1,x}^2 < \sigma_{1,y}^2$ and $\sigma_{k,x}^2 < \sigma_{k,y}^2$. In other words, interpolation decreases both the short and long variance of the series; however, the effect on the short variance, the term in the denominator, is relatively greater than on the long variance, and consequently, the overall effect is a decrease in the variance ratio.

### 6.1. The Effect of Linear Interpolation on the Volatility of Stationary Series

First, we consider the stationary series studied in previous sections.

### 6.1.1. Stationary AR(1) Process

In the case of a stationary AR(1) model, we reported in section 4.1 that the short variances of the non-interpolated and interpolated series are given, respectively, by $\sigma_{1,y}^2 = \dfrac{2}{1+\alpha}\sigma_\varepsilon^2$ and

$$\sigma_{1,x}^2 = \dfrac{2(1-\alpha^s)}{s^2(1-\alpha^2)}\sigma_\varepsilon^2.$$

The corresponding long variances of the non-interpolated and interpolated series are given, respectively, by $\sigma_{k,y}^2 = \dfrac{2(1-\alpha^s)}{1-\alpha^2}\sigma_\varepsilon^2$ and $\sigma_{k,x}^2 = \left[\dfrac{\sigma_\varepsilon^2}{3s^2(1-\alpha^2)}\right]\left[(1-\alpha^s)(3s^2+\alpha^s s^2-\alpha^s+3)\right]$.

Comparing the short variances, we have

$$\sigma_{1,x}^2 = \dfrac{2(1-\alpha^s)}{s^2(1-\alpha^2)}\sigma_\varepsilon^2 = \sigma_{1,y}^2\left[\dfrac{1-\alpha^s}{s^2(1-\alpha)}\right] < \sigma_{1,y}^2$$

where the inequality holds because $\dfrac{1-\alpha^s}{s^2(1-\alpha)} < 1$ for $|\alpha|<1$ and $s \geq 2$. It follows that the short variance of the interpolated AR(1) series is smaller than the short variance of the corresponding

non-interpolated AR(1) series. Moreover, $\lim\limits_{s \to \infty} \left[ \dfrac{1-\alpha^s}{s^2(1-\alpha)} \right] = 0$, which means that as $s$ increases, the short variance of the interpolated AR(1) series shrinks. In other words, as $s$ increases, the gap between the short variances of the non-interpolated and the interpolated AR(1) series expands.

Next, comparing the long variances, we have

$$\sigma_{k,x}^2 = \left[ \dfrac{(1-\alpha^s)(3s^2 + \alpha^s s^2 - \alpha^s + 3)}{3s^2(1-\alpha^2)} \right] \sigma_\varepsilon^2 = \sigma_{k,y}^2 \dfrac{\left[s^2(3+\alpha^s) + 3 - \alpha^s\right]}{6s^2} < \sigma_{k,y}^2,$$

where the inequality holds because $\dfrac{\left[s^2(3+\alpha^s) + 3 - \alpha^s\right]}{6s^2} < 1$ for $|\alpha| < 1$ and $s \geq 2$. It follows that the long variance of the interpolated AR(1) is smaller than the long variance of the corresponding non-interpolated AR(1) series. Moreover, $\lim\limits_{s \to \infty} \left\{ \dfrac{\left[s^2(3+\alpha^s) + 3 - \alpha^s\right]}{6s^2} \right\} = \dfrac{1}{2}$, which means that as $s$ increases, the long variance of the interpolated AR(1) series approaches $\dfrac{1}{2}\sigma_{k,y}^2$. Thus, the effect of interpolation is indeed bigger on the short variance of an AR(1) series than on its long variance.

*6.1.2. Stationary MA(1) Process*

As we showed in section 4.3, the short variances of the non-interpolated and interpolated MA(1) series are given, respectively, by

$$\sigma_{1,y}^2 = \left[2(1+\theta^2) - 2\theta\right]\sigma_\varepsilon^2 \text{ and } \sigma_{1,x}^2 = \dfrac{2}{s^2}\left[(1+\theta^2)\right]\sigma_\varepsilon^2$$

The corresponding long variances of non-interpolated and interpolated MA(1) series are given, respectively, by $\sigma_{k,y}^2 = \left[2(1+\theta^2)\right]\sigma_\varepsilon^2$ and $\sigma_{k,x}^2 = \left(\dfrac{s^2+1}{s^2}\right)(1+\theta^2)\sigma_\varepsilon^2$

Comparing the short variances, we find that

$$\sigma_{1,x}^2 = \dfrac{2}{s^2}\left[(1+\theta^2)\right]\sigma_\varepsilon^2 = \sigma_{1,y}^2 \left[\dfrac{1+\theta^2}{s^2(1+\theta^2 - \theta)}\right] < \sigma_{1,y}^2$$

where the inequality holds because $\dfrac{1+\theta^2}{s^2(1+\theta^2 - \theta)} < 1$ for $|\theta| < 1$ and $s \geq 2$. It follows that the short variance of the interpolated MA(1) series is smaller than the long variance of the corresponding

non-interpolated series. Moreover, $\lim_{s \to \infty} \left[ \dfrac{1+\theta^2}{s^2 \left(1+\theta^2 - \theta\right)} \right] = 0$, which means that as $s$ increases, the short variance of the interpolated MA(1) series decreases. In other words, as $s$ increases, the gap between the short variances of the non-interpolated and the interpolated MA(1) series increases.

Next, comparing the long variances, we find that

$$\sigma_{k,x}^2 = \left(\frac{s^2+1}{s^2}\right)\left(1+\theta^2\right)\sigma_\varepsilon^2 = \sigma_{k,y}^2 \left(\frac{s^2+1}{2s^2}\right) < \sigma_{k,y}^2$$

where the inequality holds because $\dfrac{s^2+1}{2s^2} < 1$ for $s \geq 2$. It follows that the long variance of the interpolated MA(1) series is smaller than the long variance of the corresponding non-interpolated MA(1) series. Moreover, $\lim_{s \to \infty} \left(\dfrac{s^2+1}{2s^2}\right) = \dfrac{1}{2}$, which means that as $s$ increases, the long variance of the interpolated MA(1) series approaches $\dfrac{1}{2} \sigma_{k,y}^2$. This demonstrates that the effect of interpolation on the short variance of an MA(1) series is indeed bigger in relative terms than on the long variance.

### 6.1.3. Stationary ARMA(1,1) Process

As we showed in section 4.2, the short variances of the non-interpolated and interpolated stationary ARMA(1,1) process are given, respectively, by

$$\sigma_{1,y}^2 = \left[\frac{(\alpha-1)^2 \left(1+\theta^2+2\alpha\theta\right)}{1-\alpha^2} + \left(1+\theta^2\right) + 2(\alpha-1)\theta\right]\sigma_\varepsilon^2 \text{ and}$$

$$\sigma_{1,x}^2 = \frac{2\left[\left(1-\alpha^s\right)\left(1+\alpha\theta+\theta^2\right)+\alpha\theta\left(1-\alpha^{s-2}\right)\right]}{s^2(1-\alpha^2)} \sigma_\varepsilon^2$$

The corresponding long variances of the non-interpolated and interpolated stationary ARMA(1,1) process are given, respectively, by

$$\sigma_{k,y}^2 = \left[\frac{(\alpha^s-1)^2 \left(1+\theta^2+2\alpha\theta\right)}{1-\alpha^2} + \left(1+\theta^2+2\alpha\theta\right)\left(\frac{1-\alpha^{2s}}{1-\alpha^2}\right) - 2\alpha^{s-1}\theta\right]\sigma_\varepsilon^2$$

and

$$\sigma_{k,x}^2 = \left(\frac{2s^2+1}{3}\right)\left[\frac{2\left[(1-\alpha^s)(1+\alpha\theta+\theta^2)+\alpha\theta(1-\alpha^{s-2})\right]}{s^2(1-\alpha^2)}\right]\sigma_\varepsilon^2$$

$$+\left(\frac{s^2-1}{3s^2}\right)\left[\frac{(2\alpha^{s-1}-\alpha^{2s-1})(1+\alpha\theta)(\alpha+\theta)}{1-\alpha^2}-\left(\frac{1+2\alpha\theta+\theta^2}{1-\alpha^2}\right)\right]\sigma_\varepsilon^2$$

Here, it is impossible to compare the variances analytically because of the complexity of the expressions. We have therefore compared the short and long variances numerically for $|\alpha|<1$, $|\theta|<1$, and $s = 2, 3, 4, 5, 10, 15, 20,$ and $25$. As expected, we find that $\sigma_{1,x}^2 < \sigma_{1,y}^2$ and $\sigma_{k,x}^2 < \sigma_{k,y}^2$, with properties similar to the other DGPs we study.

### *6.2. The Effect of Linear Interpolation on the Volatility of Nonstationary Series*

Next, we examine the difference-stationary series commonly used in macroeconomics applications.

### *6.2.1. Difference Stationary Process with i.i.d. Errors*

As we showed in section 5.1, the short variances of the non-interpolated and interpolated nonstationary series are given, respectively, by $\sigma_{1,y}^2 = \sigma_\varepsilon^2$ and $\sigma_{1,x}^2 = \frac{1}{s}\sigma_\varepsilon^2$.

The corresponding long variances of the non-interpolated and interpolated nonstationary series are given, respectively, by $\sigma_{k,y}^2 = s\sigma_\varepsilon^2$ and $\sigma_{k,x}^2 = \left(\frac{2s^2+1}{3s}\right)\sigma_\varepsilon^2$

Thus, comparing the short variances of the non-interpolated and interpolated nonstationary series, we find that

$$\sigma_{1,x}^2 = \frac{1}{s}\sigma_\varepsilon^2 = \sigma_{1,y}^2 \frac{1}{s} < \sigma_{1,y}^2,$$

where the inequality holds because $\frac{1}{s}<1$ for $s \geq 2$. It follows that the short variance of the interpolated nonstationary series is smaller than the short variance of the corresponding non-interpolated series. Moreover, $\lim_{s\to\infty}\left[\frac{1}{s}\right] = 0$, which means that as $s$ increases, the short variance of the interpolated nonstationary series gets smaller and smaller. In other words, as $s$ increases, the gap between the short variances of the non-interpolated and the interpolated nonstationary series gets larger and larger.

Next, comparing the long variances of the non-interpolated and interpolated nonstationary series, we find that

$$\sigma_{k,x}^2 = \left(\frac{2s^2+1}{3s}\right)\sigma_\varepsilon^2 = \sigma_{k,y}^2\left(\frac{2s^2+1}{3s^2}\right) < \sigma_{k,y}^2$$

where the inequality holds because $\frac{2s^2+1}{3s^2} < 1$ for $s \geq 2$. It follows that the long variance of the interpolated nonstationary series is smaller than the long variance of the corresponding non-interpolated nonstationary series. Moreover, $\lim_{s \to \infty}\left(\frac{2s^2+1}{3s^2}\right) = \frac{2}{3}$, which means that as $s$ increases, the long variance of the interpolated nonstationary series approaches $\frac{2}{3}\sigma_{k,y}^2$. This demonstrates that the effect of interpolation on the short variance of a nonstationary series is indeed bigger in relative terms than on the long variance.

*6.2.2. Nonstationary Process with ARMA(1,1) Errors*

As we showed in section 5.2, the short variances of the non-interpolated and interpolated nonstationary process with ARMA(1,1) errors are given, respectively, by

$$\sigma_{1,y}^2 = \frac{1+\theta^2+2\alpha\theta}{1-\alpha^2}\sigma_\varepsilon^2$$

and

$$\sigma_{1,x}^2 = \left(\frac{1}{s}\right)^2\left[s\left[\frac{1+\theta^2+2\alpha\theta}{1-\alpha^2}\right] + 2\left[\frac{(1+\alpha\theta)(\alpha+\theta)}{1-\alpha^2}\right]\sum_{j=1}^{s-1}(s-j)\alpha^{j-1}\right]\sigma_\varepsilon^2$$

The corresponding long variances of the non-interpolated and interpolated nonstationary process with ARMA(1,1) errors are given, respectively, by

$$\sigma_{k,y}^2 = s\gamma_0 + 2\sum_{j=1}^{s-1}(s-j)\gamma_j$$

where $\gamma_0 = \sigma_{1,y}^2 = \frac{1+\theta^2+2\alpha\theta}{1-\alpha^2}\sigma_\varepsilon^2$, $\gamma_1 = \frac{(1+\alpha\theta)(\alpha+\theta)}{1-\alpha^2}\sigma_\varepsilon^2$, and $\gamma_j = \alpha\gamma_{j-1}$, $j \geq 2$, and

$$\sigma_{k,x}^2 = \left(\frac{2s^2+1}{3s^2}\right)\left[s\left[\frac{1+\theta^2+2\alpha\theta}{1-\alpha^2}\sigma_\varepsilon^2\right] + 2\left[\frac{(1+\alpha\theta)(\alpha+\theta)}{1-\alpha^2}\sigma_\varepsilon^2\right]\sum_{j=1}^{s-1}(s-j)\alpha^{j-1}\right]$$
$$+ \left(\frac{s^2-1}{3s^2}\right)\left[\sum_{j=1}^{s-1}(s-j)\gamma_{s-j} + \sum_{j=0}^{s-1}(s-j)\gamma_{s+j}\right]$$

Here also, it is impossible to compare the variances analytically because of the complexity of the expressions. We have therefore compared the short and long variances numerically for $|\alpha|<1$, $|\theta|<1$, and $s = 2, 3, 4, 5, 10, 15, 20,$ and $25$. As expected, we find that $\sigma_{1,x}^2 < \sigma_{1,y}^2$ and $\sigma_{k,x}^2 < \sigma_{k,y}^2$, with properties similar to the other DGP's we study.

One systematic finding we are reporting in section 6.1 is that $\sigma_{1,x}^2 < \sigma_{1,y}^2$, which means that the short variance of the interpolated series is always smaller than the short variance of the original, non-interpolated series, and that is irrespective of the DGP we employ, and irrespective of whether the process is stationary or nonstationary.

Short variance is a standard volatility measure of a time series. Our findings, therefore, indicate that linear interpolation reduces the volatility of a time series, and that seems to be independent of the true stationarity properties of the original series. To the extent that linear interpolation was employed in constructing the prewar series, the consensus view that prewar macroeconomic series were more volatile than the postwar series because of the interpolation used in constructing the prewar series is not in line with our findings. Our findings suggest the opposite: linear interpolation would make the prewar series more stable (under the assumption that the "true" prewar and postwar series followed approximately the same DGP).

## *7*. Implications for the Prewar vs. Postwar Output Shock Persistence and Volatility Debate

Macroeconomists have long been studying the difference between the prewar and postwar output and other aggregate series. Until the 1980s, there was a consensus that the postwar US output and other aggregate series were less volatile than the corresponding prewar series (e.g., Burns 1960, Lucas 1977, Gordon 1986, DeLong and Summers 1986 and 1988, and Backus and Kehoe 1992). The change has been attributed to the success of the postwar macroeconomic stabilization policies. In fact, according to Balke and Gordon (1989, pp. 38–39), "Until recently, one of the least controversial stylized facts in macroeconomic history was the reduced volatility of output in the United States after World War II. Indeed, Arthur Burns [1960] devoted his entire 1959 American Economic Association presidential address to explaining the phenomenon of a more stable postwar economy."[16]

Romer (1999, p. 23), however, argues that real macroeconomic indicators have not really become more stable between the pre-World War I and post-World War II eras, and recessions have become only slightly less severe on average. In her view, the conclusion of a more stable postwar economy is an artifact of measurement pitfalls in the prewar data, as many prewar series were constructed using various interpolation schemes. The accuracy of the views about prewar business cycles, therefore, depends critically on the impact of interpolation on prewar macroeconomic series.

---

[16] To emphasize this consensus, Gordon (1986) notes a 1969 conference volume titled "Is Business Cycle Obsolete? (Bronfenbrenner 1969) and cites Samuelson (from Zarnowitz 1972), who said that "the NBER has worked itself out of one of its first jobs, namely, the business cycle."

In a similar vein, some empirical studies report that there are differences in the persistence properties of macroeconomic time series. These studies find that the US postwar output series exhibits higher shock persistence than the prewar series (e.g., Stock and Watson 1986, DeLong and Summers 1986, and Campbell and Mankiw 1987a, 1987b). Does this reflect a real shift in the US macroeconomic characteristics, or is this also an artifact of the measurement of prewar data?

It may be reasonable to argue that linear interpolation of the US prewar macroeconomic series might have lowered the shock persistence of these series, contributing to the difference in shock persistence of the US prewar and postwar series. However, the persuasiveness of this argument is weakened by the fact that many international prewar series, which were also interpolated, do not exhibit the low shock persistence observed for the US data. For example, comparing the persistence of output series for various countries in the prewar and postwar era, DeLong and Summers (1988) report that the U.S. prewar GNP exhibits much lower shock persistence than comparable prewar European output series like Sweden's and the United Kingdom's. Maddison (1982, Appendix A) refers to country-specific sources for the series that pertain to Australia, Canada, Denmark, France, Italy, Norway, Sweden, and the UK. As noted, our examination of these sources reveals that linear interpolation was used in building historical series of output for most of these countries, including the UK. This can be viewed as indirect evidence against the proposition that linear interpolation lowers the shock persistence of a series, and thus it is the reason for the US prewar-postwar macroeconomic differences.[17]

Our results indicate that linear interpolation, in fact, *reduces* volatility and *increases* the shock persistence of the series generated by parsimonious time series models characterizing macroeconomic data. These findings are contrary to the prevailing notion that interpolation of the US prewar data is a cause of its higher volatility and lower shock persistence when compared to US postwar data. Interestingly, our findings are consistent with the results that international prewar macroeconomic series that have been interpolated do not exhibit low shock persistence.

---

[17] Sheffrin (1988, p. 81), in a rare disagreement with the consensus view, argues that "… the heavy reliance on methods of interpolation … [for constructing the prewar series, is unlikely to] bias the data in favor of excessive volatility for the earlier periods."

Our finding's surprising implications is that the gap in the volatility and shock persistence of the prewar and postwar series would have been even larger than documented in the absence of linear interpolation. Therefore, the true prewar output series likely was even more volatile and less shock persistent than the existing literature recognizes, which is in line with the conclusion of Balke and Gordon (1989). In that case, the postwar period dampening in the business cycles may have been even a greater success story of the postwar economic policy making than it was recognized by Burns (1960) and others.

Our results are obviously subject to the same assumption-related caveats as any other analytical examination. Notwithstanding this rhetorical caution, our findings have implications for future studies. More specifically, since the low shock persistence and high volatility documented for the prewar US macroeconomic series do not seem to be an artifact of linear interpolation, what should they be attributable to? Other prewar, postwar differences in policy, institutional factors, war-driven influences, or data collection practices? This is an open question that merits further investigation. The many differences between prewar and postwar economies provide fertile ground for such investigation. For example, one explanation for the differences between the US prewar and postwar output might be a change in output composition after the war, as government spending, which exhibits more persistence, became a large component of the output. A decomposition of the income side might also point to other factors contributing to the difference between the prewar and postwar macroeconomic cycles.

Moreover, providing convincing empirical evidence on the difference in prewar and postwar economies requires using data that is not linearly interpolated and are, therefore, more reliable for such purposes. For example, Cogley and Sargent (2015) and Cogley, Sargent, and Surico (2015) overcome the difficulties posed by the inaccuracies found in the prewar data by using methods that account for measurement errors. Franses (2021) offers a method for estimating the shock persistence of irregularly spaced time series. Balke and Gordon (1989) employ previously unused output component series to generate a new output series for the 1869–1928 period. Romer (1989) relies on the actual time-varying relationship between GNP and commodity output to generate a new output series for the 1869–1908 period, while Miron and Romer (1990) use component series for generating a new monthly index of industrial production for the 1884–1940 period.

Identification of the factors that may contribute to such differences, however, can also be done indirectly by focusing on postwar data that is more reliable and drawing parallels with the prewar era. This can benefit from the approach that Stock and Watson (2003, 2007) use to study the postwar trends in US GDP and inflation data and report a reduction in the volatility of GDP and ease of predictability of inflation over time. They explore possible causes such as changes in monetary policy conduct, frequency of external shocks to productivity and oil or commodity prices, financial market frictions, changes in the structure of the real economy, such as relative growth of various sectors with different levels of stability, etc.

## 8. Concluding Remarks

Many studies have argued that interpolation can distort the statistical properties of time series (e.g., Stock and Watson 1986, Jaeger 1990, Dezhbakhsh and Levy 1994, Cheung and Chinn 1997, Murray and Nelson 2000, Charles and Darné 2010, among others). In fact, some authors suspect that the difference in statistical properties of the macroeconomic series before WW I and after WW II might be an artifact of linear interpolation of prewar data. For example, DeLong and Summers (1986), Campbell and Mankiw (1987a and 1987b), and others conjecture that the differences found between prewar and postwar shock persistence may be due to the interpolation procedure used to construct the prewar series. Moreover, some authors suspect that interpolation can potentially taint macroeconomic findings (Ehrmann 2000, Franses et al. 2006, Davis 2006, Franses 2013, Charles, et al. 2014, Williamson 2018, and Kaufmann 2020). Douglas (1930), Hanes (2006), and Keating and Valcarcel (2015) emphasize other shortcomings of linear interpolation, noting that the approach is logically improbable and easily refutable when contrasted with the observed data and that the final result is data with inferior quality.

Despite such criticisms, linearly interpolated data form the basis of much of our historical macroeconomic understanding. As argued by Romer in a series of papers, our knowledge of the prewar behavior of output, prices, employment, and, more generally, business cycles is based on these interpolated historical time series.

The analytical results presented in this paper indicate that linear interpolation of time series data leads to (a) a decrease in the volatility, and to (b) an increase in shock persistence of the series with variance ratios that are larger than the benchmark value irrespective of the parameters of the data generating process and irrespective of the interpolated segment size (number of the missing observations interpolated). All our findings hold regardless of whether the original series is a random walk with i.i.d. errors, a random walk with ARMA (1,1) errors, or a stationary AR(1), MA(1), or ARMA(1,1) process, which are the parsimonious models commonly used to characterize macroeconomic time series.

Our results also show that linear interpolation introduces periodic variation in the *k*-period growth variance of the interpolated series. This finding is in line with Romer's (1986a, 1986b, 1986c, 1986d, 1989) observations about the amplified cyclical volatility in prewar data. She explores the accuracy of these historical data (which, as she notes, economists have rarely done) in great detail and concludes that the methods used to generate much of the prewar series have amplified the cyclical volatility in the prewar series, mistakenly leading to a conclusion that the postwar data exhibits less volatility in comparison to the prewar period.

The most important implication of our finding is that alternative explanations for the difference between the shock persistence and volatility of the US prewar and postwar macro data—e.g., institutional changes, policy variables, or other macro factors—need to be explored to explain the difference in shock persistence and volatility behavior before and after the war. Studies like Basu and Taylor (1999), Blanchard and Watson (1986), Stock and Watson (2003, 2007), and Dennis et al. (2007) can perhaps provide a road map for identifying temporal drivers of such changes. Other possible avenues include evolving institutional structures and policy practices, which can potentially alter the short-run and long-run volatility and shock persistence properties of output and other macroeconomic variables (Acemoglu et al. 2003, 2005, and 2021, Acemoglu and Robinson 2008 and 2012, and Acemoglu 2025).

We end the paper with a cautionary note. Our goal in this paper was to examine the effect of linear interpolation on time series properties because of the widespread use of linear interpolation in pre-war macroeconomic data construction and the importance of such data in our understanding of macroeconomic trends before and after the war. We believe our approach offers a pathway to investigate the impact of other interpolation methods in other settings, given that the use of interpolation goes far beyond macroeconomics or even economics, with interpolation being an important data generation tool in other disciplines as well.

**References**

Acemoglu, D. (2025), "Nobel Lecture: Institutions, Technology, and Prosperity," *American Economic Review* 115(6), 1709–1748.

Acemoglu, D., Johnson, S., Robinson, J, and Thaicharoen, Y. (2003), "Institutional Causes, Macroeconomic Symptoms: Volatility, Crises, and Growth," *Journal of Monetary Economics* 50(1), 49–123.

Acemoglu, D., Johnson, S., and Robinson, J. (2005), "Institutions as a Fundamental Cause of Long-Run Growth," in *Handbook of Economic Growth, Vol. 1A* (New York, NY: Elsevier), 385–472.

Acemoglu, D., and Robinson, J. (2008), "The Persistence and Change of Institutions in the Americas," *Southern Economic Journal* 75(2), 281–299.

Acemoglu, D., and Robinson, J. (2012), *Why Nations Fail: The Origins of Power, Prosperity, and Poverty* (New York, NY: Crown Business).

Acemoglu, D., Egorov, G., and Sonin, K. (2021), "Institutional Change and Institutional Persistence," *in The Handbook of Historical Economics*, edited by A. Bisin and G. Federico (New York, NY: Academic Press), pp. 365–389.

Adorf, H. (1995), "Interpolation of Irregularly Sampled Data Series – a Survey," in: Shaw, R.A., et al. (Eds.), *Astronomical Data Analysis Software and Systems IV*, 77, 460–463.

Backus, D., and Kehoe, P. (1992), "International Evidence on the Historical Properties of Business Cycles," *American Economic Review* 82(4), 864–888.

Bailey, M. (1978), "Stabilization Policy and Private Economic Behavior," *Brookings Papers on Economic Activity* 78, 11–60.

Balke, N., and Gordon, R. (1986), "Appendix B: Historical Data." In: Gordon, R., Ed., *The American Business Cycle: Continuity and Change* (Chicago, IL: The University of Chicago Press and NBER), pp. 781–850.

Balke, N., and Gordon, R. (1989), "The Estimation of Prewar Gross National Product: Methodology and New Evidence," *Journal of Political Economy* 97(1), 38–92.

Bank of England (2018), "Notes on Three Centuries of UK GDP and 7 Centuries of English GDP.".

Basu, S., and Taylor, A. (1999), "Business Cycles in International Historical Perspective," *Journal of Economic Perspectives* 13(2), 45–68.

Bergman, M., Bordo, M., and Jonung, L. (1998), "Historical Evidence on Business Cycles: The International Experience," in: Federal Reserve Bank of Boston Conference Series, Volume 42, *Beyond Shocks: What Causes Business Cycles?* edited by J. Fuhrer and S. Schuh, pp. 65–113.

Blanchard, O., and Watson, M. (1986), "Are Business Cycles All Alike?" In: Gordon, R., Ed., *The American Business Cycle: Continuity and Change* (University of Chicago Press), 123–180.

Bronfenbrenner, M. (1969), *Is the Business Cycle Obsolete?* Proceedings of the Conference of the Social Science Research Council Committee on Economic Stability (New York, NY: Wiley).

Bureau of Labor Statistics (1966), "The Consumer Price Index: History and Techniques," *Bulletin No. 1517*, Washington, DC: U.S. Government Printing Office.

Burns, A. (1960) "Progress towards Economic Stability" *American Economic Review* 50, 1–19.

Campbell, J., and Mankiw, N. (1987a), "Are Output Fluctuations Transitory?" *Quarterly Journal of Economics* 102, 857–880.

Campbell, J. and Mankiw, N. (1987b), "Permanent and Transitory Components in Macroeconomic Fluctuations," *American Economic Review* 77, 111–117.

Campbell, G., Quinn, W., Turner, J. D., and Ye, Q. (2018a), "What Moved Share Prices in the Nineteenth-Century London Stock Market?" *Economic History Review* 71(1), 157–189.

Campbell, G., Turner, J. D., and Ye, Q. (2018b), "The Liquidity of the London Capital Markets, 1825–70," *Economic History Review* 71(3), 823–852.

Carleton, W., Campbell, D., and Collard, M. (2014), "A Reassessment of the Impact of Drought

Cycles on the Classic Maya," *Quaternary Science Reviews* 105, 151–161.

Cecchetti, S., Lam, P. (1994), "Variance-Ratio Tests: Small-Sample Properties with an Application to International Output Data," *Journal of Business and Economic Statistics* 12(2), 177–186.

Charles, A., and Darné, O. (2012), "A Note on the Uncertain Trend in US Real GNP: Evidence from Robust Unit Root Tests," *Economics Bulletin* 32(3), 2399–2406.

Charles, A., Darné, O., and Diebolt, C. (2014), "A Revision of the US Business-Cycles Chronology, 1790–1928," *Economics Bulletin* 34(1), 234–244.

Cheung Y, and Chinn, M. (1997), "Further Investigation of the Uncertain Unit Root in GNP," *Journal of Business and Economic Statistics* 15(1), 68–73.

Cochrane J. (1988), "How Big is the Random Walk in GNP?" *Journal of Political Economy* 96, 893–820.

Cogley T. (1990), "International Evidence on the Size of the Random Walk in Output," *Journal of Political Economy* 98(3), 501–518.

Cogley, T., and Sargent, T. (2015), "Measuring Price-Level Uncertainty and Instability in the United States, 1850–2012," *Review of Economics and Statistics* 97(4), 827–838.

Cogley, T., Sargent, T., and Surico, P. (2015), "Price-Level Uncertainty and Instability in the United Kingdom," *Journal of Economic Dynamics and Control* 52, 1–16.

Davis, J. (2006), "An Improved Annual Chronology of U.S. Business Cycles Since the 1790s," *Journal of Economic History* 66, 103–121.

Dennis, B., and Işcan, T. (2007), "Accounting for Structural Change: Evidence from Two Centuries of U.S. Data," Working Paper No. 2007-04, Dept. of Econ., Dalhousie University.

DeLong, B., and Summers, L. (1986), "The Changing Cyclical Variability of Economic Activity in the United States," in: *The American Business Cycle: Continuity and Change*, edited by R. Gordon (Chicago, IL: University of Chicago Press), pp. 679–734.

DeLong, B., and Summers, L. (1988), "How Does Macroeconomic Policy Affect Output?" *Brookings Papers on Economic Activity* 2, 433–480.

Dezhbakhsh, H., and Levy, D. (1994), "Periodic Properties of Interpolated Time Series," *Economics Letters* 44, 221–228.

Dezhbakhsh, H., and Levy, D. (2003), "International Evidence on Output Fluctuation and Shock Persistence," *Journal of Monetary Economics* 50, 1531–1553.

Dezhbakhsh, H., and Levy, D. (2022), "Interpolation and Shock Persistence of Prewar U.S. Macroeconomic Time Series: A Reconsideration," *Economics Letters* 213, 110386,

Douglas, P. (1930), *Real Wages in the United States, 1890–1926* (Boston, MA: Houghton).

Ehrmann, M. (2000), "Comparing Monetary Policy Transmission across European Countries," *Weltwirtschaftliches Archiv* 136(1), 58–83.

Fohlin, C., and Collet, S. (2025), "Hedging Against Turmoil: Asset Pricing, Liquidity, and Equity Market Dynamics in the German Hyperinflation," manuscript, Dept. of Econ., Emory University.

Franses, P., Paap, R., and Fok, D. (2006), "Performance of Seasonal Adjustment Procedures: Simulation and Empirical Results," in T. Mills and K. Patterson (eds.), *Palgrave Handbook of Econometrics*, *Volume I: Econometric Theory* (Palgrave MacMillan, New York), pp. 1035–1055.

Franses, P. (2013), "Data Revisions and Periodic Properties of Macroeconomic Data," *Economics Letters* 120, 139–141.

Franses, P. (2021), "Estimating Persistence for Irregularly Spaced Historical Data," *Quality & Quantity* 55, 2177–2187.

Franses, P. (2022), "Interpolation and Correlation," *Applied Economics* 54(14), 1562–1567.

Friedman, M., and Schwartz, A. (1982), *Monetary Trends in the US and the UK* (NBER and the University of Chicago Press).

Fuhrer, J., and Schuh, S. (1998), "Beyond Shocks: What Causes Business Cycles? An Overview," in: Federal Reserve Bank of Boston Conference Series, Volume 42, *Beyond Shocks: What*

*Causes Business Cycles?* edited by J. Fuhrer and S. Schuh, pp. 1–31.

Gordon, R. (1986), "Introduction: Continuity and Change in Theory, Behavior, and Methodology," in: *The American Business Cycle: Continuity and Change*, ed. by R. Gordon (Chicago, IL), 1–34.

Jaeger, A. (1990), "Shock Persistence and the Measurement of Prewar Output Series," *Economics Letters* 34, 333–337.

Jane, N., Nehemiah, K., and Arputharaj, K. (2016), "A Temporal Mining Framework for Classifying Unevenly Spaced Clinical Data," *Applied Clinical Informatics* 7, 1–21.

Johansson, O. (1967), *The Gross Domestic Product of Sweden and Its Composition 1861–1955* (Stockholm: Almqvist and Wiksell).

Johnston, L., and Williamson, S. (2018), "The Annual Real and Nominal GDP for the United States, 1790–1928," www.measuringworth.com/datasets/usgdp12/sourcegdp.php.

Karger, E. and Wray, A. (2024), "The Black-White Lifetime Earnings Gap," *Explorations in Economic History* 94, 101629.

Katz, J., and Sanger-Katz, M. (2020), "Coronavirus Deaths by US State and Country over Time: Daily Tracker," *New York Times*, March 28, 2020, accessed March 29, 2020.

Kaufmann, D. (2020), "Is Deflation Costly After All? The Perils of Erroneous Historical Classifications," *Journal of Applied Econometrics* 35(5), 614–628.

Keating, J. and Valcarcel, V. (2015), "The Time-Varying Effects of Permanent and Transitory Shocks to Real Output," *Macroeconomic Dynamics* 19, 477–507.

Kuznets, S. (1961), *Capital in the American Economy: Its Formation and Financing* (NBER and the University of Chicago Press).

Lebergott, S. (1964), *Manpower in Economic Growth: The American Record Since 1800* (McGraw-Hill: New York, NY).

Leung, S. (1992), "Changes in the Behavior of Output in the United Kingdom, 1856–1990," *Economics Letters* 40, 435–444.

Levy, D., and Chen, H. (1994), "Estimates of the Aggregate Quarterly Capital Stock Series for the Postwar U.S. Economy," *Review of Income and Wealth* 40, 317–349.

Levy, D., Snir, A., Chen, H., and Gotler, A. (2020), "Not All Price Endings Are Created Equal: Price Points and Asymmetric Price Rigidity," *Journal of Monetary Economics* 110, 33–49.

Liu, Z. (2016), "Time Series Modeling of Irregularly Sampled Multivariate Clinical Data," PhD Thesis, Department of Computer Science, University of Pittsburgh.

Liu, Z., and Hauskrecht, M. (2016), "Learning Adaptive Forecasting Models from Irregularly Sampled Multivariate Clinical Data," *Proceedings of the Thirtieth AAAI Conference on Artificial Intelligence (AAAI-16)*, 1273–1279.

Long, C. (1960), *Wages and Earnings in the United States, 1860–1890* (Princeton, NJ: Princeton University Press and NBER).

Lucas, R. E., Jr. (1977), "Understanding Business Cycles," *Carnegie-Rochester Conference Series on Public Policy* 5, 7–29.

Ma, Y., de Jong, H., and Chu, T. (2014), "Living Standards in China between 1840 and 1912: a New Estimate of Gross Domestic Product per Capita," GGDC Research Memorandum; Vol. GD-147), Groningen Growth and Development Center.

Maddison, A. (1982), *Phases of Capitalist Development* (New York, NY: Oxford University Press).

Miron, J., and Romer, C. (1990), "A New Monthly Index of Industrial Production, 1884–1940," *Journal of Economic History* 50(2), 321–337.

Murray, C., and Nelson, C. (2000), "The Uncertain Trend in U.S. GDP," *Journal of Monetary Economics* 46(1), 79–95.

Orlando, O., and Zimatore, G. (2018), "Recurrence Quantification Analysis of Business Cycles," *Chaos, Solitons and Fractals* 110, 82–94.

Romer, C. (1986a), "Is the Stabilization of the Postwar Economy a Figment of the Data?" *American Economic Review* 76(3), 314–334.

Romer, C. (1986b), "The Instability of the Prewar Economy Reconsidered: A Critical Examination of Historical Macroeconomic Data," *Journal of Economic History* 46(2), 494–496.

Romer, C. (1986c), "New Estimates of Prewar Gross National Product and Unemployment," *Journal of Economic History* 46(2), 341–352.

Romer, C. (1986d), "Spurious Volatility in Historical Unemployment Data," *Journal of Political Economy* 94(1), l–37.

Romer, C. (1989), "The Prewar Business Cycle Reconsidered: New Estimates of Gross National Product, 1869–1908," *Journal of Political Economy* 97, 1–37.

Romer, C. (1991), "The Cyclical Behavior of Individual Production Series, 1889-1984," *Quarterly Journal of Economics* 106(1), 1–31.

Romer, C. (1992), "What Ended the Great Depression?" *Journal of Economic History* 52(4), 757–784.

Romer, C. (1994), "Remeasuring Business Cycles," *Journal of Economic History* 54(3), 573–609.

Romer, C. (1999), "Changes in Business Cycles: Evidence and Explanations," *Journal of Economic Perspectives* 13(2), 23–44.

Ramey, G., and Ramey, V. (1995), "Cross-Country Evidence on the Link between Volatility and Growth," *American Economic Review* 85(5), 1138–1151.

Rousseau, P. L. (2009), "Share Liquidity, Participation, and Growth of the Boston Market for Industrial Equities, 1854–1897," *Explorations in Economic History* 46(2), 203–219.

Samuelson, P. (1998), "Summing Up on Business Cycles: Opening Address," in: Federal Reserve Bank of Boston Conference Series, Volume 42, *Beyond Shocks: What Causes Business Cycles?* edited by J. Fuhrer and S. Schuh, pp. 33–36.

Sheffrin, S. (1988), "Have Economic Fluctuations Been Dampened? A Look at the Evidence Outside the United States," *Journal of Monetary Economics* 21, 73–83.

Shi, M. (2015), "Forest Cover Change in Northeast China during the Period of 1977–2007 and Its Driving Forces," PhD Thesis, Michigan State University.

Shi, M., R. Yin, and Lv, H. (2017), "An Empirical Analysis of the Driving Forces of Forest Cover Change in Northeast China," *Forest Policy and Economics* 78, 78–87.

Stock, J. (1994), "Unit Roots, Structural Breaks, and Trends," in *Handbook of Econometrics*, Vol. IV, Edited by R. Engle and D. McFadden (New York, NY: Elsevier), 2739–2841.

Stock, J., & Watson, M. (1986) "Does GNP Have a Unit Root?" *Economic Letters* 22, 147–51.

Stock, J., and Watson, M. (2003), "Has the Business Cycle Changed and Why?" *NBER Macroeconomics Annual* 17, 159–218.

Stock, J., and Watson, M. (2007), "Why Has US Inflation Become Harder to Forecast," *Journal of Money, Credit, and Banking* 30 (1), 3–33.

Taylor, J. (1986), "Improvements in Macroeconomic Stability: The Role of Wages and Prices," in R. Gordon, ed., *The American Business Cycle: Continuity and Change* (Chicago, IL: University of Chicago Press), pp. 639–659.

Tobin, J. (1980), *Asset Accumulation and Economic Activity* (Chicago, IL: Univ. of Chicago Press).

Williamson, S. (2018), "Measuring Worth Is Better Without the CPI," manuscript, presented at the American Economic Association annual conference.

Zarnowitz, V. (1972), *The Business Cycle Today* (New York, NY: Columbia University Press).

Zarnowitz, V. (1992), *Business Cycles: Theory, History, Indicators, and Forecasting* (Chicago, IL: University of Chicago Press).

Zarnowitz, V. (1998), "Has the Business Cycle Been Abolished," *Business Economics* 33, 39–45.

Zimatore, O. (2020) "Business Cycle Modeling between Financial Crises and Black Swans: Ornstein-Uhlenbeck Stochastic Process vs Kaldor Deterministic Chaotic Model," *Chaos* 30(8), 083129.

**Figure 1. Variance Ratio $V_y(\alpha, \theta)$ for ARMA(1, 1) Series**

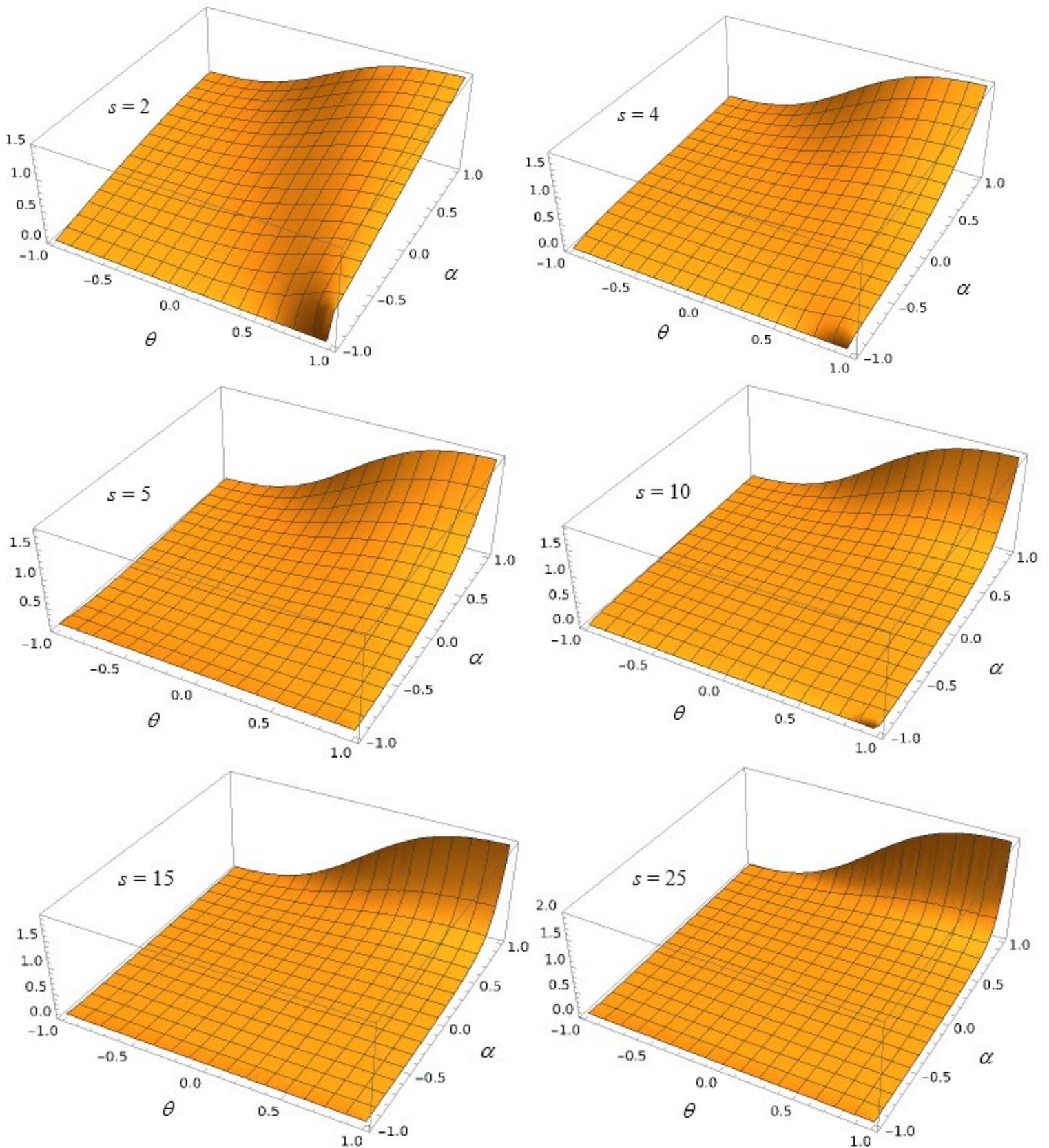

**Figure 2. Variance Ratio $V_x(\alpha, \theta)$ for the Interpolated ARMA(1, 1) Series**

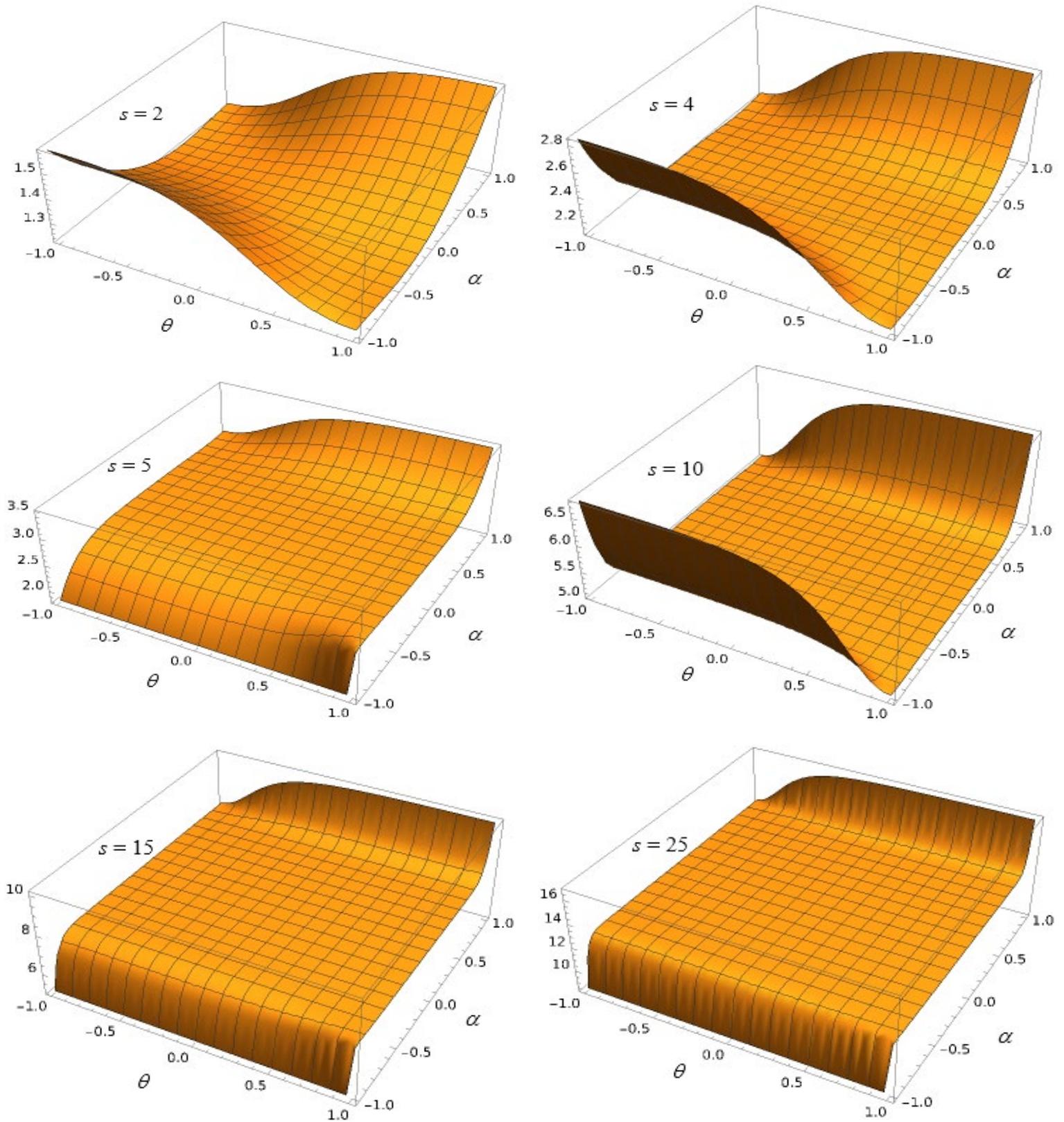

**Figure 3. Variance Ratio $V_y(\alpha, \theta)$ for Nonstationary Series with ARMA(1, 1) Errors**

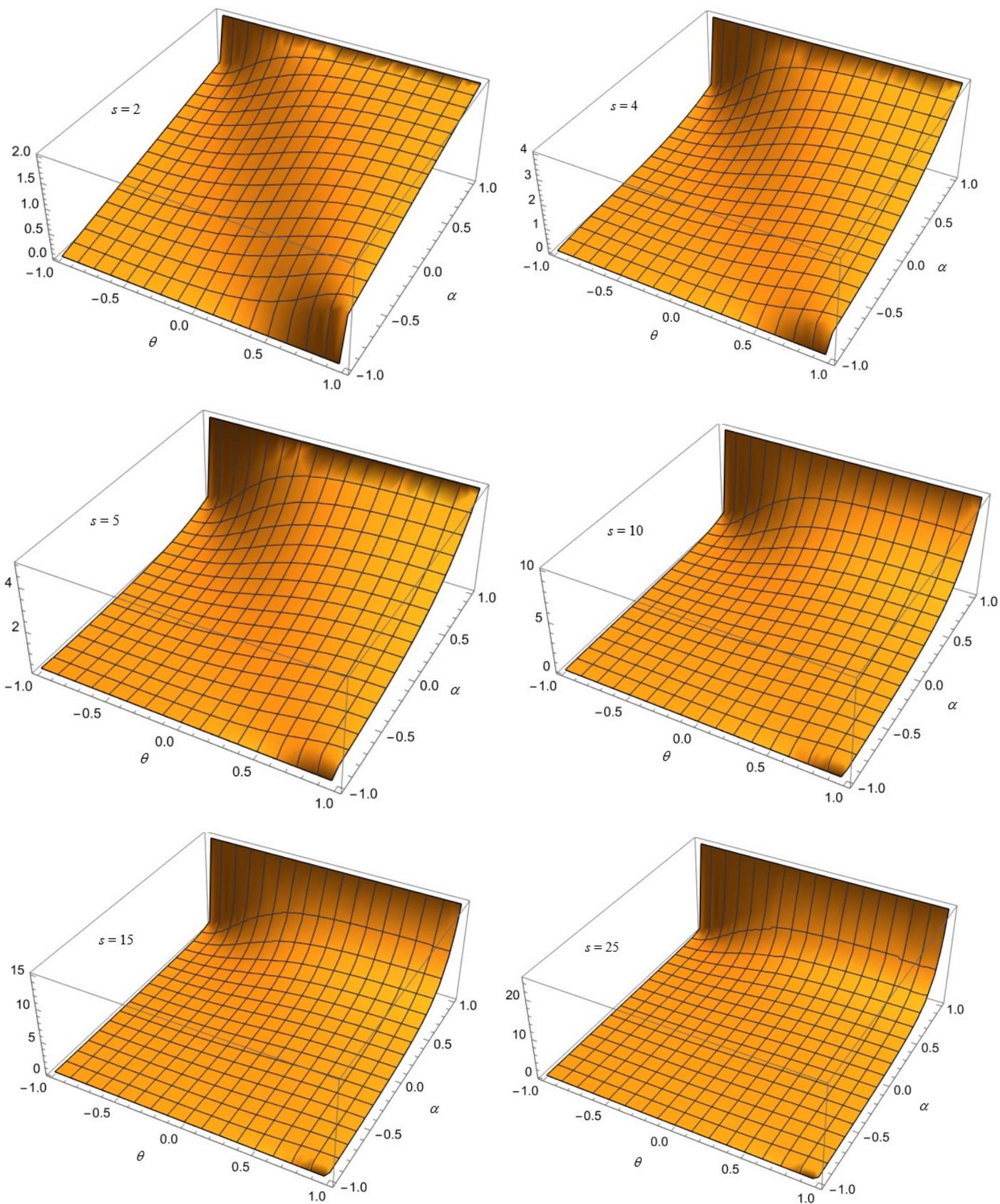

**Figure 4. Variance Ratio $V_x(\alpha, \theta)$ for the Interpolated Nonstationary Series with ARMA(1, 1) Errors**

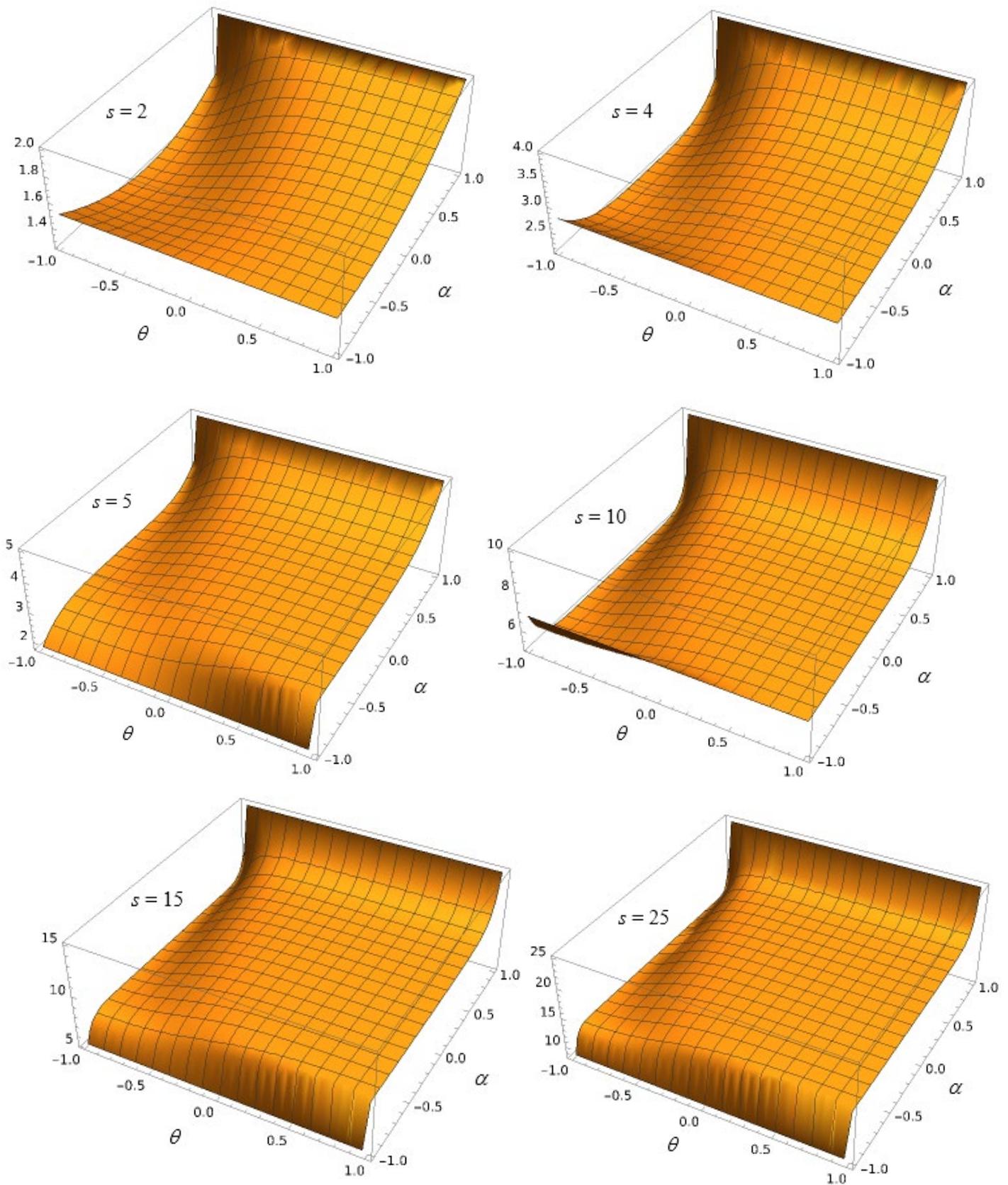

**Table 1. Variance Ratio of a Stationary ARMA(1, 1) Model, for the Original Series $V_y$, and the Interpolated Series $V_x$, for $s = 4$**

| $\alpha$ | $\theta$ | 0.99 | 0.90 | 0.75 | 0.50 | 0.10 | 0.00 | –0.10 | –0.50 | –0.75 | –0.90 | –0.99 |
|---|---|---|---|---|---|---|---|---|---|---|---|---|
| 0.99 | $V_y$ | 1.73 | 1.72 | 1.70 | 1.58 | 1.13 | 0,98 | 0.84 | 0.40 | 0.28 | 0.25 | 0.25 |
|  | $V_x$ | 2.81 | 2.81 | 2.81 | 2.80 | 2.75 | 2.72 | 2.69 | 2.42 | 2.21 | 2.14 | 2.12 |
| 0.90 | $V_y$ | 1.54 | 1.53 | 1.51 | 1.40 | 0.99 | 0.86 | 0.73 | 0.36 | 0.27 | 0.25 | 0.25 |
|  | $V_x$ | 2.61 | 2.61 | 2.61 | 2.60 | 2.56 | 2.53 | 2.51 | 2.31 | 2.17 | 2.12 | 2.12 |
| 0.75 | $V_y$ | 1.26 | 1.25 | 1.23 | 1.13 | 0.79 | 0.68 | 0.58 | 0.31 | 0.25 | 0.24 | 0.23 |
|  | $V_x$ | 2.38 | 2.37 | 2.37 | 2.37 | 2.34 | 2.32 | 2.30 | 2.19 | 2.12 | 2.11 | 2.10 |
| 0.50 | $V_y$ | 0.91 | 0.90 | 0.88 | 0.80 | 0.54 | 0.47 | 0.41 | 0.25 | 0.21 | 0.21 | 0.21 |
|  | $V_x$ | 2.18 | 2.18 | 2.18 | 2.18 | 2.17 | 2.16 | 2.16 | 2.12 | 2.11 | 2.10 | 2.10 |
| 0.10 | $V_y$ | 0.55 | 0.55 | 0.53 | 0.47 | 0.31 | 0.28 | 0.25 | 0.19 | 0.17 | 0.17 | 0.17 |
|  | $V_x$ | 2.12 | 2.12 | 2.12 | 2.12 | 2.12 | 2.12 | 2.12 | 2.12 | 2.12 | 2.12 | 2.12 |
| 0.00 | $V_y$ | 0.50 | 0.50 | 0.48 | 0.42 | 0.28 | NA | 0.23 | 0.18 | 0.17 | 0.17 | 0.17 |
|  | $V_x$ | 2.12 | 2.12 | 2.12 | 2.12 | 2.12 | NA | 2.12 | 2.12 | 2.12 | 2.12 | 2.12 |
| –0.10 | $V_y$ | 0.45 | 0.45 | 0.43 | 0.37 | 0.25 | 0.23 | 0.21 | 0.17 | 0.16 | 0.16 | 0.16 |
|  | $V_x$ | 2.12 | 2.12 | 2.12 | 2.12 | 2.12 | 2.12 | 2.12 | 2.12 | 2.12 | 2.12 | 2.12 |
| –0.50 | $V_y$ | 0.34 | 0.34 | 0.32 | 0.25 | 0.17 | 0.16 | 0.15 | 0.13 | 0.13 | 0.13 | 0.13 |
|  | $V_x$ | 2.11 | 2.11 | 2.11 | 2.12 | 2.16 | 2.16 | 2.17 | 2.18 | 2.18 | 2.18 | 2.18 |
| –0.75 | $V_y$ | 0.30 | 0.29 | 0.25 | 0.16 | 0.10 | 0.10 | 0.09 | 0.08 | 0.08 | 0.08 | 0.08 |
|  | $V_x$ | 2.10 | 2.11 | 2.12 | 2.19 | 2.30 | 2.32 | 2.34 | 2.37 | 2.37 | 2.37 | 2.37 |
| –0.90 | $V_y$ | 0.27 | 0.25 | 0.17 | 0.08 | 0.05 | 0.04 | 0.04 | 0.04 | 0.04 | 0.04 | 0.04 |
|  | $V_x$ | 2.12 | 2.12 | 2.17 | 2.31 | 2.51 | 2.53 | 2.56 | 2.60 | 2.61 | 2.61 | 2.61 |
| –0.99 | $V_y$ | 0.25 | 0.12 | 0.03 | 0.01 | 0.00 | 0.00 | 0.00 | 0.00 | 0.00 | 0.00 | 0.00 |
|  | $V_x$ | 2.12 | 2.14 | 2.21 | 2.42 | 2.69 | 2.72 | 2.75 | 2.80 | 2.81 | 2.81 | 2.81 |

Variance Ratio of the Original (Non-Interpolated) Series:

$$V_y = \left(\frac{1}{s}\right) \left[\frac{(1+\theta^2+2\alpha\theta) - \alpha^{s-1}(1+\alpha\theta)(\alpha+\theta)}{(1+\theta^2+2\alpha\theta) - (1+\alpha\theta)(\alpha+\theta)}\right]$$

Variance Ratio of the Interpolated Series:

$$V_x = \left(\frac{2s^2+1}{3s}\right) + \left(\frac{s^2-1}{6s}\right) \left[\frac{\alpha^{s-1}(2-\alpha^s)(1+\alpha\theta)(\alpha+\theta) - (1+2\alpha\theta+\theta^2)}{(1-\alpha^s)(1+2\alpha\theta+\theta^2) + \alpha\theta(1-\alpha^{s-2})}\right]$$

**Table 2. Variance Ratio for a Stationary ARMA(1, 1) Model, for the Original Series $V_y$, and the Interpolated Series $V_x$, for $s = 10$**

| $\alpha$ | $\theta$ | 0.99 | 0.90 | 0.75 | 0.50 | 0.10 | 0.00 | –0.10 | –0.50 | –0.75 | –0.90 | –0.99 |
|---|---|---|---|---|---|---|---|---|---|---|---|---|
| 0.99 | $V_y$ | 1.82 | 1.82 | 1.79 | 1.65 | 1.13 | 0.96 | 0.78 | 0.27 | 0.13 | 0.10 | 0.10 |
|  | $V_x$ | 6.62 | 6.62 | 6.62 | 6.61 | 6.57 | 6.54 | 6.50 | 6.10 | 5.47 | 5.12 | 5.05 |
| 0.90 | $V_y$ | 1.26 | 1.26 | 1.24 | 1.14 | 0.77 | 0.65 | 0.54 | 0.20 | 0.12 | 0.10 | 0.10 |
|  | $V_x$ | 5.67 | 5.67 | 5.67 | 5.67 | 5.64 | 5.62 | 5.60 | 5.39 | 5.15 | 5.05 | 5.03 |
| 0.75 | $V_y$ | 0.75 | 0.74 | 0.73 | 0.66 | 0.44 | 0.38 | 0.31 | 0.14 | 0.10 | 0.09 | 0.09 |
|  | $V_x$ | 5.16 | 5.16 | 5.16 | 5.16 | 5.15 | 5.14 | 5.14 | 5.09 | 5.05 | 5.04 | 5.04 |
| 0.50 | $V_y$ | 0.40 | 0.40 | 0.39 | 0.35 | 0.23 | 0.20 | 0.17 | 0.10 | 0.08 | 0.08 | 0.08 |
|  | $V_x$ | 5.05 | 5.05 | 5.05 | 5.05 | 5.05 | 5.05 | 5.05 | 5.05 | 5.05 | 5.05 | 5.05 |
| 0.10 | $V_y$ | 0.22 | 0.22 | 0.21 | 0.18 | 0.12 | 0.11 | 0.10 | 0.07 | 0.07 | 0.07 | 0.07 |
|  | $V_x$ | 5.05 | 5.05 | 5.05 | 5.08 | 5.05 | 5.05 | 5.05 | 5.05 | 5.05 | 5.05 | 5.05 |
| 0.00 | $V_y$ | 0.20 | 0.20 | 0.19 | 0.17 | 0.11 | N/A | 0.09 | 0.07 | 0.07 | 0.07 | 0.07 |
|  | $V_x$ | 5.05 | 5.05 | 5.05 | 5.05 | 5.05 | N/A | 5.05 | 5.05 | 5.05 | 5.05 | 5.05 |
| –0.10 | $V_y$ | 0.18 | 0.18 | 0.17 | 0.15 | 0.10 | 0.09 | 0.08 | 0.07 | 0.06 | 0.06 | 0.06 |
|  | $V_x$ | 5.05 | 5.05 | 5.05 | 5.05 | 5.05 | 5.05 | 5.05 | 5.05 | 5.05 | 5.05 | 5.05 |
| –0.50 | $V_y$ | 0.13 | 0.13 | 0.12 | 0.10 | 0.07 | 0.07 | 0.06 | 0.06 | 0.06 | 0.06 | 0.06 |
|  | $V_x$ | 5.05 | 5.05 | 5.05 | 5.05 | 5.05 | 5.05 | 5.05 | 5.05 | 5.05 | 5.05 | 5.05 |
| –0.75 | $V_y$ | 0.11 | 0.11 | 0.10 | 0.07 | 0.05 | 0.05 | 0.05 | 0.05 | 0.05 | 0.05 | 0.05 |
|  | $V_x$ | 5.04 | 5.04 | 5.05 | 5.09 | 5.14 | 5.14 | 5.15 | 5.16 | 5.16 | 5.16 | 5.16 |
| –0.90 | $V_y$ | 0.11 | 0.10 | 0.07 | 0.05 | 0.03 | 0.03 | 0.03 | 0.03 | 0.03 | 0.03 | 0.03 |
|  | $V_x$ | 5.03 | 5.05 | 5.15 | 5.39 | 5.60 | 5.62 | 5.64 | 5.67 | 5.67 | 5.67 | 5.67 |
| –0.99 | $V_y$ | 0.10 | 0.05 | 0.01 | 0.01 | 0.00 | 0.00 | 0.00 | 0.00 | 0.00 | 0.00 | 0.00 |
|  | $V_x$ | 5.05 | 5.12 | 5.47 | 6.10 | 6.50 | 6.54 | 6.57 | 6.61 | 6.62 | 6.62 | 6.62 |

Variance Ratio of the Original (Non-Interpolated) Series:

$$V_y = \left(\frac{1}{s}\right) \left[ \frac{(1+\theta^2+2\alpha\theta) - \alpha^{s-1}(1+\alpha\theta)(\alpha+\theta)}{(1+\theta^2+2\alpha\theta) - (1+\alpha\theta)(\alpha+\theta)} \right]$$

Variance Ratio of the Interpolated Series:

$$V_x = \left(\frac{2s^2+1}{3s}\right) + \left(\frac{s^2-1}{6s}\right) \left[ \frac{\alpha^{s-1}(2-\alpha^s)(1+\alpha\theta)(\alpha+\theta) - (1+2\alpha\theta+\theta^2)}{(1-\alpha^s)(1+2\alpha\theta+\theta^2) + \alpha\theta(1-\alpha^{s-2})} \right]$$

**Table 3. Variance Ratio of the Nonstationary Model with ARMA(1, 1) Errors, for the Original Series $V_y$, and the Interpolated Series $V_x$, for $s = 4$**

| $\alpha$ | $\theta$ | 0.99 | 0.90 | 0.75 | 0.50 | 0.10 | 0.00 | –0.10 | –0.50 | –0.75 | –0.90 | –0.99 |
|---|---|---|---|---|---|---|---|---|---|---|---|---|
| 0.99 | $V_y$ | 3.96 | 3.96 | 3.96 | 3.96 | 3.95 | 3.95 | 3.94 | 3.84 | 3.38 | 2.04 | 1.00 |
|  | $V_x$ | 3.97 | 3.97 | 3.97 | 3.97 | 3.97 | 3.96 | 3.96 | 3.95 | 3.89 | 3.58 | 2.75 |
| 0.90 | $V_y$ | 3.66 | 3.66 | 3.66 | 3.65 | 3.57 | 3.52 | 3.46 | 2.76 | 1.64 | 1.00 | 0.86 |
|  | $V_x$ | 3.71 | 3.71 | 3.71 | 3.71 | 3.70 | 3.69 | 3.69 | 3.59 | 3.26 | 2.75 | 2.54 |
| 0.75 | $V_y$ | 3.21 | 3.21 | 3.21 | 3.17 | 2.99 | 2.90 | 2.77 | 1.79 | 1.00 | 0.73 | 0.68 |
|  | $V_x$ | 3.38 | 3.38 | 3.38 | 3.38 | 3.36 | 3.35 | 3.34 | 3.16 | 2.75 | 2.41 | 2.32 |
| 0.50 | $V_y$ | 2.59 | 2.59 | 2.58 | 2.52 | 2.21 | 2.06 | 1.89 | 1.00 | 0.59 | 0.49 | 0.47 |
|  | $V_x$ | 3.07 | 3.07 | 3.07 | 3.06 | 3.03 | 3.02 | 2.99 | 2.75 | 2.39 | 2.20 | 2.16 |
| 0.10 | $V_y$ | 1.88 | 1.88 | 1.85 | 1.75 | 1.31 | 1.16 | 1.00 | 0.47 | 0.32 | 0.28 | 0.28 |
|  | $V_x$ | 2.86 | 2.86 | 2.86 | 2.85 | 2.81 | 2.12 | 2.75 | 2.48 | 2.23 | 2.14 | 2.12 |
| 0.00 | $V_y$ | 1.75 | 1.74 | 1.72 | 1.60 | 1.15 | 1.00 | 0.85 | 0.40 | 0.28 | 0.25 | 0.25 |
|  | $V_x$ | 2.84 | 2.84 | 2.84 | 2.83 | 2.78 | 2.75 | 2.71 | 2.44 | 2.21 | 2.14 | 2.12 |
| –0.10 | $V_y$ | 1.63 | 1.63 | 1.60 | 1.46 | 1.00 | 0.86 | 0.72 | 0.34 | 0.25 | 0.23 | 0.23 |
|  | $V_x$ | 2.82 | 2.82 | 2.82 | 2.81 | 2.75 | 2.72 | 2.68 | 2.40 | 2.20 | 2.14 | 2.12 |
| –0.50 | $V_y$ | 1.28 | 1.27 | 1.22 | 1.00 | 0.53 | 0.44 | 0.36 | 0.20 | 0.16 | 0.16 | 0.16 |
|  | $V_x$ | 2.77 | 2.77 | 2.77 | 2.75 | 2.65 | 2.61 | 2.56 | 2.31 | 2.20 | 2.17 | 2.16 |
| –0.75 | $V_y$ | 1.13 | 1.11 | 1.00 | 0.68 | 0.28 | 0.23 | 0.19 | 0.11 | 0.10 | 0.10 | 0.10 |
|  | $V_x$ | 2.75 | 2.75 | 2.75 | 2.73 | 2.63 | 2.59 | 2.55 | 2.39 | 2.34 | 2.32 | 2.32 |
| –0.90 | $V_y$ | 1.05 | 1.00 | 0.77 | 0.37 | 0.12 | 0.09 | 0.08 | 0.05 | 0.05 | 0.04 | 0.04 |
|  | $V_x$ | 2.75 | 2.75 | 2.75 | 2.73 | 2.67 | 2.65 | 2.63 | 2.56 | 2.54 | 2.53 | 2.53 |
| –0.99 | $V_y$ | 1.00 | 0.64 | 0.20 | 0.05 | 0.01 | 0.01 | 0.01 | 0.00 | 0.00 | 0.00 | 0.00 |
|  | $V_x$ | 2.75 | 2.75 | 2.75 | 2.75 | 2.74 | 2.74 | 2.73 | 2.73 | 2.72 | 2.72 | 2.72 |

Variance Ratio of the Original (Non-Interpolated) Series:

$$V_y = 1 + \frac{2}{s}\left[\frac{(1+\alpha\theta)(\alpha+\theta)}{1+\theta^2+2\alpha\theta}\right]\left[\frac{s(1-\alpha)-(1-\alpha^s)}{(1-\alpha)^2}\right]$$

Variance Ratio of the Interpolated Series:

$$V_x = \left(\frac{2s^2+1}{3s}\right) + \frac{(1+\alpha\theta)(\alpha+\theta)(s^2-1)(1-\alpha^s)^2}{3s^2(1-\alpha)^2(1+\theta^2+2\alpha\theta)+6s(1+\alpha\theta)(\alpha+\theta)\left[s(1-\alpha)-(1-\alpha^s)\right]}$$

**Table 4. Variance Ratio of the Nonstationary Model with ARMA(1, 1) Errors, for the Original Series $V_y$, and the Interpolated Series $V_x$, for $s=10$**

| $\alpha$ | $\theta$ | 0.99 | 0.90 | 0.75 | 0.50 | 0.10 | 0.00 | −0.10 | −0.50 | −0.75 | −0.90 | −0.99 |
|---|---|---|---|---|---|---|---|---|---|---|---|---|
| 0.99 | $V_y$ | 9.72 | 9.72 | 9.72 | 9.71 | 9.69 | 9.68 | 9.65 | 9.34 | 7.99 | 4.07 | 1.00 |
|      | $V_x$ | 9.79 | 9.79 | 9.79 | 9.79 | 9.79 | 9.79 | 9.79 | 9.77 | 9.71 | 9.30 | 6.70 |
| 0.90 | $V_y$ | 7.62 | 7.62 | 7.62 | 7.58 | 7.39 | 7.28 | 7.12 | 5.38 | 2.60 | 1.00 | 0.65 |
|      | $V_x$ | 8.44 | 8.44 | 8.44 | 8.44 | 8.43 | 8.43 | 8.42 | 8.33 | 7.93 | 6.70 | 5.64 |
| 0.75 | $V_y$ | 5.36 | 5.35 | 5.34 | 5.28 | 4.92 | 4.73 | 4.48 | 2.56 | 1.00 | 0.47 | 0.38 |
|      | $V_x$ | 7.47 | 7.47 | 7.47 | 7.46 | 7.45 | 7.44 | 7.43 | 7.27 | 6.70 | 5.64 | 5.15 |
| 0.50 | $V_y$ | 3.40 | 3.40 | 3.38 | 3.29 | 2.82 | 2.60 | 2.34 | 1.00 | 0.38 | 0.23 | 0.20 |
|      | $V_x$ | 6.99 | 6.99 | 6.99 | 6.99 | 6.96 | 6.95 | 6.93 | 6.70 | 6.04 | 5.29 | 5.05 |
| 0.10 | $V_y$ | 2.09 | 2.08 | 2.05 | 1.92 | 1.39 | 1.20 | 1.00 | 0.35 | 0.16 | 0.12 | 0.11 |
|      | $V_x$ | 6.81 | 6.81 | 6.80 | 6.80 | 6.76 | 6.73 | 6.70 | 6.31 | 5.61 | 5.15 | 5.05 |
| 0.00 | $V_y$ | 1.90 | 1.89 | 1.86 | 1.72 | 1.18 | 1.00 | 0.82 | 0.28 | 0.14 | 0.10 | 0.10 |
|      | $V_x$ | 6.79 | 6.79 | 6.78 | 6.78 | 6.73 | 6.70 | 6.66 | 6.23 | 5.53 | 5.14 | 5.05 |
| −0.10 | $V_y$ | 1.74 | 1.74 | 1.70 | 1.55 | 1.00 | 0.83 | 0.67 | 0.23 | 0.12 | 0.09 | 0.09 |
|       | $V_x$ | 6.77 | 6.77 | 6.77 | 6.76 | 6.70 | 6.67 | 6.62 | 6.14 | 5.47 | 5.12 | 5.05 |
| −0.50 | $V_y$ | 1.31 | 1.30 | 1.24 | 1.00 | 0.48 | 0.38 | 0.29 | 0.11 | 0.07 | 0.07 | 0.07 |
|       | $V_x$ | 6.73 | 6.73 | 6.72 | 6.70 | 6.57 | 6.51 | 6.42 | 5.76 | 5.25 | 5.08 | 5.05 |
| −0.75 | $V_y$ | 1.13 | 1.11 | 1.00 | 0.66 | 0.24 | 0.19 | 0.15 | 0.07 | 0.06 | 0.05 | 0.05 |
|       | $V_x$ | 6.71 | 6.71 | 6.70 | 6.65 | 6.42 | 6.32 | 6.19 | 5.53 | 5.23 | 5.15 | 5.14 |
| −0.90 | $V_y$ | 1.05 | 1.00 | 0.77 | 0.36 | 0.11 | 0.08 | 0.07 | 0.04 | 0.03 | 0.03 | 0.03 |
|       | $V_x$ | 6.70 | 6.70 | 6.69 | 6.63 | 6.39 | 6.29 | 6.18 | 5.79 | 5.66 | 5.63 | 5.62 |
| −0.99 | $V_y$ | 1.00 | 0.65 | 0.20 | 0.05 | 0.01 | 0.01 | 0.01 | 0.00 | 0.00 | 0.00 | 0.00 |
|       | $V_x$ | 6.70 | 6.70 | 6.70 | 6.68 | 6.64 | 6.62 | 6.61 | 6.56 | 6.54 | 6.54 | 6.54 |

Variance Ratio of the Original (Non-Interpolated) Series:

$$V_y = 1 + \frac{2}{s}\left[\frac{(1+\alpha\theta)(\alpha+\theta)}{1+\theta^2+2\alpha\theta}\right]\left[\frac{s(1-\alpha)-(1-\alpha^s)}{(1-\alpha)^2}\right]$$

Variance Ratio of the Interpolated Series:

$$V_x = \left(\frac{2s^2+1}{3s}\right) + \frac{(1+\alpha\theta)(\alpha+\theta)(s^2-1)(1-\alpha^s)^2}{3s^2(1-\alpha)^2(1+\theta^2+2\alpha\theta)+6s(1+\alpha\theta)(\alpha+\theta)\left[s(1-\alpha)-(1-\alpha^s)\right]}$$

# Interpolation and Prewar-Postwar Output Volatility and Shock-Persistence Debate: A Closer Look and New Results*

**Hashem Dezhbakhsh**
Department of Economics, Emory University
Atlanta, GA 30322, USA
econhd@emory.edu

**Daniel Levy**
Department of Economics, Bar-Ilan University
Ramat-Gan 5290002, Israel
Department of Economics, Emory University
Atlanta, GA 30322, USA
ICEA, ISET at TSU, and RCEA
Daniel.Levy@biu.ac.il

February 10, 2026

## Appendix A. Derivations for the Stationary Process

### A-I. Variance Ratio for AR(1) Series is Smaller than 1

**Lemma:** We shall prove that $V_y < 1$, by proving that $V_y^{-1} > 1$ by induction. Thus, we need to prove that $V_y^{-1} = \frac{s(1-\alpha)}{1-\alpha^s} > 1$ for $s \geq 2$ and $|\alpha| < 1$.

**Proof:** Let $s = 2$. Then $\frac{s(1-\alpha)}{1-\alpha^s} = \frac{2(1-\alpha)}{1-\alpha^2} > 1$, which implies that $\alpha^2 - 2\alpha + 1 > 0$, which is equivalent to $(\alpha - 1)^2 > 0$. The latter inequality holds for all $|\alpha| < 1$. Next assume that $\frac{s(1-\alpha)}{1-\alpha^s} > 1$ and prove for $s+1$: $\frac{(s+1)(1-\alpha)}{1-\alpha^{s+1}} > 1 \Rightarrow (s+1)(1-\alpha) > (1-\alpha^{s+1}) \Rightarrow s(1-\alpha) > \alpha(1-\alpha^s) \Rightarrow \frac{s(1-\alpha)}{1-\alpha^s} > \alpha$, which is satisfied because $\frac{s(1-\alpha)}{1-\alpha^s} > 1 > \alpha$.

### Q.E.D

**********************************************************************

### A-II. Derivations for the Long Variance of Interpolated AR(1) Series

Following the notation in section 3.1,

$$\sigma_{k,x}^2 = \text{var}(x_{t,i} - x_{t-1,i})$$

$$= i^2 \text{var}\left[\frac{y_{t,s} - y_{t-1,s}}{s}\right] + (s-i)^2 \text{var}\left[\frac{y_{t-1,s} - y_{t-2,s}}{s}\right] + 2i(s-i)\text{cov}\left[\left(\frac{y_{t,s} - y_{t-1,s}}{s}\right), \left(\frac{y_{t-1,s} - y_{t-2,s}}{s}\right)\right]$$

$$= i^2 \left[\frac{2(1-\alpha^s)}{s^2(1-\alpha^2)}\sigma_\varepsilon^2\right] + (s-i)^2 \left[\frac{2(1-\alpha^s)}{s^2(1-\alpha^2)}\sigma_\varepsilon^2\right]$$

$$+ \frac{2i(s-i)}{s^2}\left[\text{cov}(y_{t,s}, y_{t-1,s}) - \text{cov}(y_{t,s}, y_{t-2,s}) - \text{cov}(y_{t-1,s}, y_{t-1,s}) + \text{cov}(y_{t-1,s}, y_{t-2,s})\right]$$

$$= \left[i^2 + (s-i)^2\right]\left[\frac{2(1-\alpha^s)}{s^2(1-\alpha^2)}\sigma_\varepsilon^2\right] + \left[\frac{2i(s-i)}{s^2}\right](\gamma_s - \gamma_{2s} - \gamma_0 + \gamma_s)$$

where $\gamma_0 = \frac{\sigma_\varepsilon^2}{1-\alpha^2}$, and $\gamma_j = \alpha^j \frac{\sigma_\varepsilon^2}{1-\alpha^2} = \alpha^j \gamma_0$ for $j \geq 1$ are the covariance functions of the AR(1) series. Further simplification of the long variance expression leads to

$$\sigma_{k,x}^2 = (s^2 - 2si + 2i^2)\left[\frac{2(1-\alpha^s)}{s^2(1-\alpha^2)}\sigma_\varepsilon^2\right] + \left[\frac{2si - 2i^2}{s^2}\right](2\alpha^s - \alpha^{2s} - 1)\left(\frac{\sigma_\varepsilon^2}{1-\alpha^2}\right),$$

where we note that the *k*-period growth variance of the interpolated series depends on *i*, implying that it is characterized by *periodic variation*. This is a recurring property in the results we report in the paper. This point is consistent with the finding that linearly interpolated series exhibit periodic variation regardless of the true nature of the original, non-interpolated series.

To eliminate the dependence of the variance on $i$, we use the expected value of $i$ in the variance expression.[1]

$$\sigma_{k,x}^2 = \left[ s^2 - 2s \frac{\left(\sum_{i=1}^{s} i\right)}{s} + 2 \frac{\left(\sum_{i=1}^{s} i^2\right)}{s} \right] \left[ \frac{2(1-\alpha^s)}{s^2(1-\alpha^2)} \sigma_\varepsilon^2 \right] + \left(\frac{1}{s^2}\right) \left[ 2s \frac{\left(\sum_{i=1}^{s} i\right)}{s} - 2 \frac{\left(\sum_{i=1}^{s} i^2\right)}{s} \right] (2\alpha^s - \alpha^{2s} - 1) \left(\frac{\sigma_\varepsilon^2}{1-\alpha^2}\right)$$

$$= \left[ s^2 - \left(\frac{2s}{s}\right) \frac{s(s+1)}{2} + \left(\frac{2}{s}\right) \frac{s(s+1)(2s+1)}{6} \right] \left[ \frac{2(1-\alpha^s)}{s^2(1-\alpha^2)} \sigma_\varepsilon^2 \right]$$

$$+ \left(\frac{1}{s^2}\right) \left[ \left(\frac{2s}{s}\right) \frac{s(s+1)}{2} - \left(\frac{2}{s}\right) \frac{s(s+1)(2s+1)}{6} \right] (2\alpha^s - \alpha^{2s} - 1) \left(\frac{\sigma_\varepsilon^2}{1-\alpha^2}\right)$$

$$= \left[ s^2 - s(s+1) + \frac{(s+1)(2s+1)}{3} \right] \left[ \frac{2(1-\alpha^s)}{s^2(1-\alpha^2)} \sigma_\varepsilon^2 \right]$$

$$+ \left(\frac{1}{s^2}\right) \left[ s(s+1) - \frac{(s+1)(2s+1)}{3} \right] (2\alpha^s - \alpha^{2s} - 1) \left(\frac{\sigma_\varepsilon^2}{1-\alpha^2}\right)$$

$$= \left[ \frac{(2s^2+1)}{3} \right] \left[ \frac{2(1-\alpha^s)}{s^2(1-\alpha^2)} \sigma_\varepsilon^2 \right] + \left(\frac{1}{s^2}\right) \left[ \frac{(s^2-1)}{3} \right] (2\alpha^s - \alpha^{2s} - 1) \left(\frac{\sigma_\varepsilon^2}{1-\alpha^2}\right)$$

$$= \left[ \frac{\sigma_\varepsilon^2}{3s^2(1-\alpha^2)} \right] \left[ 2(2s^2+1)(1-\alpha^s) + (s^2-1)(2\alpha^s - \alpha^{2s} - 1) \right]$$

$$= \left[ \frac{\sigma_\varepsilon^2}{3s^2(1-\alpha^2)} \right] \left[ (1-\alpha^s)(3s^2 + \alpha^s s^2 - \alpha^s + 3) \right]$$

\*\*\*\*\*\*\*\*\*\*\*\*\*\*\*\*\*\*\*\*\*\*\*\*\*\*\*\*\*\*\*\*\*\*\*\*\*\*\*\*\*\*\*\*\*\*\*\*\*\*\*\*\*\*\*\*\*\*\*\*\*\*\*\*

## A-III. Variance Ratio for a Linearly Interpolated AR(1) Series is Larger than One

**Lemma:** $V_x = \dfrac{s^2(3+\alpha^s) + (3-\alpha^s)}{6s} > 1$ for $s \geq 2$ and $|\alpha| < 1$.

---

[1] If we have series that consist of equal number of observations from $\{x_i\}$ and $\{y_i\}$ with $\text{var}(x_i) = \sigma_x^2$ and $\text{var}(y_i) = \sigma_y^2$, where both $\{x_i\}$ and $\{y_i\}$ are zero-mean independent series, then,

$$E\left[\text{var}\left\{\begin{matrix}x_i\\y_i\end{matrix}\right\}\right] = E\left[\sum_{i=1}^{T/2} x_i^2 + \sum_{i=1}^{T/2} x_i^2\right]\frac{1}{T} = \left[\sum_{i=1}^{T/2} \sigma_x^2 + \sum_{i=1}^{T/2} \sigma_y^2\right]\frac{1}{T} = \frac{\sigma_x^2 + \sigma_y^2}{2},$$ which is the average of the two variances. In deriving $E\left[\sigma_{k,x}^2(i)\big|i\right]$ below, we have $s$ sub-series, and thus we compute the average of $s$ variances.

**Proof:** If $s=2$, then $V_x = \dfrac{s^2(3+\alpha^s)+(3-\alpha^s)}{6s} = \dfrac{4(3+\alpha^2)+(3-\alpha^2)}{12} = \dfrac{15+3\alpha^2}{12} > 1$, since $|\alpha|<1$.

Next, assume $\dfrac{s^2(3+\alpha^s)+(3-\alpha^s)}{6s} > 1$, and prove $\dfrac{(s+1)^2(3+\alpha^{s+1})+(3-\alpha^{s+1})}{6(s+1)} > 1$.

Then $(s+1)^2(3+\alpha^{s+1})+(3-\alpha^{s+1})-6(s+1) > 0 \Rightarrow 3s^2 + s(s\alpha^{s+1}+2\alpha^{s+1}) > 0$.

However, $|\alpha|<1 \Rightarrow \alpha^{s+1} > -1 \Rightarrow 2\alpha^{s+1} > -2 \Rightarrow s\alpha^{s+1} > -s \Rightarrow 2\alpha^{s+1} + s\alpha^{s+1} > -2-s \Rightarrow$

$s(2\alpha^{s+1} + s\alpha^{s+1}) > -s(2+s) \Rightarrow 3s^2 + s(2\alpha^{s+1} + s\alpha^{s+1}) > 3s^2 - s(2+s) \Rightarrow$

$3s^2 + s(2\alpha^{s+1} + s\alpha^{s+1}) > 2s(s-1)$. However, $2s(s-1) > 0$ for $s \geq 2$.

Therefore, $3s^2 + s(2\alpha^{s+1} + s\alpha^{s+1}) > 0$.    **QED**

*********************************************

**A-IV. Derivations for the Long Variance and Variance Ratio for ARMA(1,1) Series**

The long difference for the ARMA(1,1) series can be expressed as

$$y_{t,s} - y_{t-1,s} = (\alpha^s - 1)y_{t-1,s} + \sum_{j=0}^{s-2} \alpha^j \left(\varepsilon_{t,s-j} + \theta\varepsilon_{t,s-j-1}\right) + \alpha^{s-1}\left(\varepsilon_{t,1} + \theta\varepsilon_{t-1,s}\right)$$

Therefore,

$$\sigma_{k,y}^2 = \text{var}(y_{t,s} - y_{t-1,s})$$

$$= (\alpha^s - 1)^2 \sigma_y^2 + (1+\theta^2)\sigma_\varepsilon^2 \left(\sum_{j=1}^{s} \alpha^{2(j-1)}\right) + 2\theta\sigma_\varepsilon^2 \left(\sum_{j=1}^{s-1} \alpha^j \alpha^{j-1}\right) + 2(\alpha^s - 1)\alpha^{s-1}\theta\sigma_\varepsilon^2$$

$$= (\alpha^s - 1)^2 \sigma_y^2 + (1+\theta^2)\sigma_\varepsilon^2 \left(\sum_{j=1}^{s} \alpha^{2(j-1)}\right) + 2\theta\sigma_\varepsilon^2 \left(\sum_{j=1}^{s} \alpha^j \alpha^{j-1}\right) - 2\alpha^{s-1}\theta\sigma_\varepsilon^2$$

$$= (\alpha^s - 1)^2 \sigma_y^2 + (1+\theta^2)\sigma_\varepsilon^2 \left(\sum_{j=1}^{s} \alpha^{2(j-1)}\right) + 2\alpha\theta\sigma_\varepsilon^2 \left(\sum_{j=1}^{s} \alpha^{2(j-1)}\right) - 2\alpha^{s-1}\theta\sigma_\varepsilon^2$$

$$= (\alpha^s - 1)^2 \sigma_y^2 + (1+\theta^2 + 2\alpha\theta)\sigma_\varepsilon^2 \left(\dfrac{1-\alpha^{2s}}{1-\alpha^2}\right) - 2\alpha^{s-1}\theta\sigma_\varepsilon^2$$

Again, we have used the fact that $\sum_{j=1}^{s}\alpha^{2(j-1)} = \dfrac{1-\alpha^{2s}}{1-\alpha^2}$ in the last step of the above derivation.

The variance ratio of the series, therefore, can be computed using the short variance that is $\sigma_{1,y}^2 = \text{var}(y_{t,s} - y_{t,s-1}) = (\alpha-1)^2 \sigma_y^2 + (1+\theta^2)\sigma_\varepsilon^2 + 2(\alpha-1)\theta\sigma_\varepsilon^2$, and the long variance given above. Therefore,

$$V_y = \frac{\sigma_{k,y}^2}{k\sigma_{1,y}^2}$$

$$= \left(\frac{1}{s}\right)\left[\frac{(\alpha^s-1)^2\sigma_y^2 + (1+\theta^2)\sigma_\varepsilon^2\left(\frac{1-\alpha^{2s}}{1-\alpha^2}\right) + 2\theta\sigma_\varepsilon^2\left(\frac{\alpha-\alpha^{2s+1}}{1-\alpha^2}\right) - 2\alpha^{s-1}\theta\sigma_\varepsilon^2}{(\alpha-1)^2\sigma_y^2 + (1+\theta^2)\sigma_\varepsilon^2 + 2(\alpha-1)\theta\sigma_\varepsilon^2}\right]$$

$$= \left(\frac{1}{s}\right)\left\{\frac{(\alpha^s-1)^2\left[\frac{(1+\theta^2+2\alpha\theta)}{1-\alpha^2}\sigma_\varepsilon^2\right] + (1+\theta^2)\sigma_\varepsilon^2\left(\frac{1-\alpha^{2s}}{1-\alpha^2}\right) + 2\theta\sigma_\varepsilon^2\left(\frac{\alpha-\alpha^{2s+1}}{1-\alpha^2}\right) - 2\alpha^{s-1}\theta\sigma_\varepsilon^2}{(\alpha-1)^2\left[\frac{(1+\theta^2+2\alpha\theta)}{1-\alpha^2}\sigma_\varepsilon^2\right] + (1+\theta^2)\sigma_\varepsilon^2 + 2(\alpha-1)\theta\sigma_\varepsilon^2}\right\}$$

$$= \left(\frac{1}{s}\right)\left[\frac{(\alpha^s-1)^2(1+\theta^2+2\alpha\theta) + (1+\theta^2)(1-\alpha^{2s}) + 2\alpha\theta(1-\alpha^{2s}) - 2\alpha^{s-1}\theta(1-\alpha^2)}{(\alpha-1)^2(1+\theta^2+2\alpha\theta) + (1+\theta^2)(1-\alpha^2) + 2\theta(\alpha-1)(1-\alpha^2)}\right]$$

$$= \left(\frac{1}{s}\right)\left[\frac{2(1+\theta^2+2\alpha\theta) - 2\alpha^{s-1}(1+\alpha\theta)(\alpha+\theta)}{2(1+\theta^2+2\alpha\theta) - 2(1+\alpha\theta)(\alpha+\theta)}\right]$$

(A1)
$$= \left(\frac{1}{s}\right)\left[\frac{(1+\theta^2+2\alpha\theta) - \alpha^{s-1}(1+\alpha\theta)(\alpha+\theta)}{(1+\theta^2+2\alpha\theta) - (1+\alpha\theta)(\alpha+\theta)}\right]$$

\*\*\*\*\*\*\*\*\*\*\*\*\*\*\*\*\*\*\*\*\*\*\*\*\*\*\*\*\*\*\*\*\*\*\*\*\*\*

## A-V. Assessing Variance Ratio for ARMA(1,1)

**Proposition:** Given a stationary ARMA(1, 1) process, if the first order autocorrelation of the process $\rho_1 < 0$, then $V_y < 1$. If $\rho_1 > 0$, then $V_y < 1$ as long as $\rho_1 < \frac{s-1}{s-\alpha^{s-1}}$.

**Proof:** Define $A \equiv 1+\theta^2+2\alpha\theta$ and $B \equiv (1+\alpha\theta)(\alpha+\theta)$. Then it follows from the long variance expression (A1) that

(A2) $$V_y = \left(\frac{1}{s}\right)\left[\frac{(1+\theta^2+2\alpha\theta) - \alpha^{s-1}(1+\alpha\theta)(\alpha+\theta)}{(1+\theta^2+2\alpha\theta) - (1+\alpha\theta)(\alpha+\theta)}\right] = \left(\frac{1}{s}\right)\left[\frac{A-\alpha^{s-1}B}{A-B}\right]$$

Note that $A > 0$ because

$$A = 1+\theta^2+2\alpha\theta$$
$$= 1+\theta^2+2\alpha\theta+(\alpha^2-\alpha^2)$$
$$= (\alpha+\theta)^2 + (1-\alpha^2) > 0,$$

where the inequality follows from the fact that $|\alpha|<1$ which means that $\alpha^2 < 1$. Note also that both the numerator and the denominator of the variance ratio in (A2) are positive because they are both variances. Thus, $A > \alpha^{s-1} B$ and $A > B$. Finally, note that the first-order autocorrelation of ARMA(1,1) series is given by

$$\rho_1 = \frac{(1+\alpha\theta)(\alpha+\theta)}{1+\theta^2+2\alpha\theta} = \frac{B}{A}.$$

Consider two possibilities: (i) $\rho_1 < 0$, and (ii) $\rho_1 > 0$. If $\rho_1 < 0$, then $B < 0$ because $A > 0$. Consequently, for $V_y = \left(\frac{1}{s}\right)\left[\frac{A-\alpha^{s-1}B}{A-B}\right]$ to be less than 1, we need $A - \alpha^{s-1}B < s(A-B)$, which means $B(s-\alpha^{s-1}) < A(s-1)$. We know $(s-\alpha^{s-1}) > 0$ because $|\alpha| < 1$ and since $B < 0$, it should be that $B(s-\alpha^{s-1}) < 0$. However, $s \geq 2$, which along with $A > 0$ implies that $A(s-1) > 0$. It follows, therefore, that $B(s-\alpha^{s-1}) < A(s-1)$, and thus $V_y < 1$.

On the other hand, if $\rho_1 > 0$, then $B > 0$ by the above formulae for $\rho_1$ because $A > 0$. Since $B > 0$ and $(1+\alpha\theta) > 0$, because $|\alpha\theta| < 1$, then it follows that $(\alpha+\theta) > 0$ as well. From the inequality $V_y = \left(\frac{1}{s}\right)\left[\frac{A-\alpha^{s-1}B}{A-B}\right] < 1$, we have $\frac{B}{A} < \frac{s-1}{s-\alpha^{s-1}}$. But $\rho_1 = \frac{B}{A}$. Thus, in this case, for the inequality $V_y < 1$ to hold, we need $\rho_1 < \frac{s-1}{s-\alpha^{s-1}}$. **Q.E.D**

\*\*\*\*\*\*\*\*\*\*\*\*\*\*\*\*\*\*\*\*\*\*\*\*\*\*\*\*\*\*\*\*\*\*\*\*\*\*

**A-VI. Variances and Variance Ratio for Interpolated ARMA(1,1) Series**

As can be seen from Section 4.1.2, the one-period difference for the interpolated series $x_{t,i}$ is given by $x_{t,i} - x_{t,i-1} = \frac{1}{s}(y_{t,s} - y_{t-1,s})$, where the $y$'s are the original series that follow an ARMA(1,1) and can, therefore, be expressed as

$$y_{t,s} - y_{t-1,s} = (\alpha^s - 1)y_{t-1,s} + \sum_{j=0}^{s-2} \alpha^j \left(\varepsilon_{t,s-j} + \theta\varepsilon_{t,s-j-1}\right) + \alpha^{s-1}\left(\varepsilon_{t,1} + \theta\varepsilon_{t-1,s}\right).$$

Accordingly, the one-period differenced interpolated series is

$$x_{t,i} - x_{t,i-1} = \frac{1}{s}\left[(\alpha^s - 1)y_{t-1,s} + \sum_{j=0}^{s-2} \alpha^j \left(\varepsilon_{t,s-j} + \theta\varepsilon_{t,s-j-1}\right) + \alpha^{s-1}\left(\varepsilon_{t,1} + \theta\varepsilon_{t-1,s}\right)\right],$$

and its variance, which is the short variance, can be derived as follows.

$$\sigma_{1,x}^2 = \text{var}(x_{t,i} - x_{t,i-1})$$

$$= \left(\frac{1}{s}\right)^2 \text{var}\left[(\alpha^s - 1)y_{t-1,s} + \sum_{j=0}^{s-2} \alpha^j (\varepsilon_{t,s-j} + \theta\varepsilon_{t,s-j-1}) + \alpha^{s-1}(\varepsilon_{t,1} + \theta\varepsilon_{t-1,s})\right]$$

$$= \left(\frac{1}{s}\right)^2 \left[(\alpha^s - 1)^2 \sigma_y^2 + (1+\theta^2)\sigma_\varepsilon^2 \left(\frac{1-\alpha^{2s}}{1-\alpha^2}\right) + 2\theta\sigma_\varepsilon^2 \left(\frac{\alpha - \alpha^{2s+1}}{1-\alpha^2}\right) - 2\alpha^{s-1}\theta\sigma_\varepsilon^2\right]$$

$$= \frac{\sigma_\varepsilon^2}{s^2(1-\alpha^2)}\left[(\alpha^s - 1)^2(1 + 2\alpha\theta + \theta^2) + (1+\theta^2)(1-\alpha^{2s}) + 2\theta\alpha(1-\alpha^{2s}) - 2\alpha^{s-1}\theta(1-\alpha^2)\right]$$

$$= \frac{2\sigma_\varepsilon^2}{s^2(1-\alpha^2)}\left[(1-\alpha^s)(1+\alpha\theta+\theta^2) + \alpha\theta(1-\alpha^{s-2})\right],$$

noting that $\sigma_y^2 = \frac{(1+\theta^2+2\alpha\theta)}{1-\alpha^2}\sigma_\varepsilon^2$.

The long variance can be derived by taking the variance of the $k$-period ($s$-period) differenced interpolated series $x_{t,i} - x_{t-1,i}$ which is related to the $k$-period differenced ARMA(1,1) series $y_{t,i} - y_{t-1,i}$ as follows.

$$x_{t,i} - x_{t-1,i} = \frac{i}{s}y_{t,s} + \frac{s-i}{s}y_{t-1,s} - \frac{i}{s}y_{t-1,s} - \frac{s-i}{s}y_{t-2,s} = i\left(\frac{y_{t,s} - y_{t-1,s}}{s}\right) + (s-i)\left(\frac{y_{t-1,s} - y_{t-2,s}}{s}\right)$$

Therefore, the long variance is obtained through the following derivation.

$$\sigma_{k,x}^2 = \text{var}(x_{t,i} - x_{t-1,i})$$

$$= i^2 \text{var}\left[\frac{y_{t,s} - y_{t-1,s}}{s}\right] + (s-i)^2 \text{var}\left[\frac{y_{t-1,s} - y_{t-2,s}}{s}\right]$$

$$+ 2i(s-i)\text{cov}\left[\left(\frac{y_{t,s} - y_{t-1,s}}{s}\right), \left(\frac{y_{t-1,s} - y_{t-2,s}}{s}\right)\right]$$

$$= \left[i^2 + (s-i)^2\right]\text{var}\left[\frac{y_{t,s} - y_{t-1,s}}{s}\right]$$

$$+ \frac{2i(s-i)}{s^2}\left[\text{cov}(y_{t,s}, y_{t-1,s}) - \text{cov}(y_{t,s}, y_{t-2,s}) - \text{cov}(y_{t-1,s}, y_{t-1,s}) + \text{cov}(y_{t-1,s}, y_{t-2,s})\right]$$

$$= \left[s^2 + 2i^2 - 2si\right]\left[\frac{2\sigma_\varepsilon^2}{s^2(1-\alpha^2)}\right]\left[(1-\alpha^s)(1+\alpha\theta+\theta^2) + \alpha\theta(1-\alpha^{s-2})\right]$$

$$+ \frac{2i(s-i)}{s^2}(\gamma_s - \gamma_{2s} - \gamma_0 + \gamma_s)$$

$$= \left[s^2 + 2i^2 - 2si\right]\left[\frac{2\sigma_\varepsilon^2}{s^2(1-\alpha^2)}\right]\left[(1-\alpha^s)(1+\alpha\theta+\theta^2) + \alpha\theta(1-\alpha^{s-2})\right]$$

$$+ \frac{2i(s-i)}{s^2}\left[(2\alpha^{s-1} - \alpha^{2s-1})\gamma_1 - \gamma_0\right]$$

where the autocovariances in the last two rows of the equation are given by $\gamma_0 = \frac{1+2\alpha\theta+\theta^2}{1-\alpha^2}\sigma_\varepsilon^2$,

$\gamma_1 = \dfrac{(1+\alpha\theta)(\alpha+\theta)}{1-\alpha^2}\sigma_\varepsilon^2$, and $\gamma_j = \alpha^{j-1}\gamma_1$ for $j>1$. To eliminate the dependency of this expression on $i$, we compute the expected value of the variance estimate, conditional on $i$.

$$\sigma_{k,x}^2 = \left[s^2 + 2\dfrac{\left(\sum_{i=1}^{s} i^2\right)}{s} - 2s\dfrac{\left(\sum_{i=1}^{s} i\right)}{s}\right]\left[\dfrac{2\sigma_\varepsilon^2}{s^2(1-\alpha^2)}\right]\left[(1-\alpha^s)(1+\alpha\theta+\theta^2)+\alpha\theta(1-\alpha^{s-2})\right]$$

$$+ 2\left[\dfrac{s\dfrac{\left(\sum_{i=1}^{s} i\right)}{s} - \dfrac{\left(\sum_{i=1}^{s} i^2\right)}{s}}{s^2}\right]\left[(2\alpha^{s-1}-\alpha^{2s-1})\gamma_1 - \gamma_0\right]$$

$$= \left[s^2 + \dfrac{(s+1)(2s+1)}{3} - s(s+1)\right]\left[\dfrac{2\sigma_\varepsilon^2}{s^2(1-\alpha^2)}\right]\left[(1-\alpha^s)(1+\alpha\theta+\theta^2)+\alpha\theta(1-\alpha^{s-2})\right]$$

$$+ 2\left[\dfrac{\dfrac{s(s+1)}{2} - \dfrac{(s+1)(2s+1)}{6}}{s^2}\right]\left[(2\alpha^{s-1}-\alpha^{2s-1})\gamma_1 - \gamma_0\right]$$

$$= \left(\dfrac{2s^2+1}{3}\right)\left[\dfrac{2\sigma_\varepsilon^2}{s^2(1-\alpha^2)}\right]\left[(1-\alpha^s)(1+\alpha\theta+\theta^2)+\alpha\theta(1-\alpha^{s-2})\right]$$

$$+ \left(\dfrac{s^2-1}{3s^2}\right)\left[(2\alpha^{s-1}-\alpha^{2s-1})\dfrac{(1+\alpha\theta)(\alpha+\theta)}{1-\alpha^2}\sigma_\varepsilon^2 - \left(\dfrac{1+2\alpha\theta+\theta^2}{1-\alpha^2}\right)\sigma_\varepsilon^2\right]$$

Therefore, the variance ratio statistic of the interpolated series is given by

$$V_x = \dfrac{\sigma_{k,x}^2}{k\sigma_{1,x}^2}$$

(A3)
$$= \dfrac{\left(\dfrac{2s^2+1}{3}\right)\left[\dfrac{2\sigma_\varepsilon^2}{s^2(1-\alpha^2)}\right]\left[(1-\alpha^s)(1+\alpha\theta+\theta^2)+\alpha\theta(1-\alpha^{s-2})\right]}{s\dfrac{2\sigma_\varepsilon^2}{s^2(1-\alpha^2)}\left[(1-\alpha^s)(1+\alpha\theta+\theta^2)+\alpha\theta(1-\alpha^{s-2})\right]}$$

$$+ \dfrac{\left(\dfrac{s^2-1}{3s^2}\right)\left[(2\alpha^{s-1}-\alpha^{2s-1})\dfrac{(1+\alpha\theta)(\alpha+\theta)}{1-\alpha^2}\sigma_\varepsilon^2 - \left(\dfrac{1+2\alpha\theta+\theta^2}{1-\alpha^2}\right)\sigma_\varepsilon^2\right]}{s\dfrac{2\sigma_\varepsilon^2}{s^2(1-\alpha^2)}\left[(1-\alpha^s)(1+\alpha\theta+\theta^2)+\alpha\theta(1-\alpha^{s-2})\right]}$$

$$= \left(\dfrac{2s^2+1}{3s}\right) + \left(\dfrac{s^2-1}{6s}\right)\left[\dfrac{\alpha^{s-1}(2-\alpha^s)(1+\alpha\theta)(\alpha+\theta) - (1+2\alpha\theta+\theta^2)}{(1-\alpha^s)(1+2\alpha\theta+\theta^2)+\alpha\theta(1-\alpha^{s-2})}\right]$$

## Appendix B. Derivations for the Nonstationary Process

## B-I. Variance Ratio Derivations for Nonstationary Process with ARMA(1,1) Errors

The one-period, short variance of this process is given by $\sigma_{1,y}^2 = \text{var}(y_{t,i} - y_{t,i-1}) = \text{var}(u_{t,i}) = \gamma_0$, where $\gamma_0 = \frac{1+\theta^2 + 2\alpha\theta}{1-\alpha^2}\sigma_\varepsilon^2$. The $k$-period, long, variance can be obtained as follows.

$$\sigma_{k,y}^2 = \text{var}(y_{t,i} - y_{t-1,i}) = \text{var}\left(\sum_{j=1}^{s} u_{t,j}\right)$$

$$= \left[\text{var}(u_{t,1}) + \cdots + \text{var}(u_{t,s})\right] + \left[2\text{cov}(u_{t,1}, u_{t,2}) + \cdots + 2\text{cov}(u_{t,1}, u_{t,s})\right]$$
$$+ \left[2\text{cov}(u_{t,2}, u_{t,3}) + \cdots + 2\text{cov}(u_{t,2}, u_{t,s})\right] + \cdots + 2\text{cov}(u_{t,s-1}, u_{t,s})$$

$$= \gamma_0 + \gamma_0 + \cdots + \gamma_0$$
$$+ 2\gamma_1 + 2\gamma_2 + 2\gamma_3 + 2\gamma_4 + \cdots + 2\gamma_{s-2} + 2\gamma_{s-1}$$
$$+ 2\gamma_1 + 2\gamma_2 + 2\gamma_3 + 2\gamma_4 + \cdots + 2\gamma_{s-2}$$
$$\vdots$$
$$+ 2\gamma_1 + 2\gamma_2 + 2\gamma_3 + 2\gamma_4$$
$$+ 2\gamma_1 + 2\gamma_2 + 2\gamma_3$$
$$+ 2\gamma_1 + 2\gamma_2$$
$$+ 2\gamma_1$$

$$= s\gamma_0 + 2(s-1)\gamma_1 + 2(s-2)\gamma_2 + 2(s-3)\gamma_3 + \cdots + 2\gamma_{s-1}$$

$$= s\gamma_0 + 2\sum_{j=1}^{s-1}(s-j)\gamma_j$$

where $\gamma_1 = \frac{(1+\alpha\theta)(\alpha+\theta)}{1-\alpha^2}\sigma_\varepsilon^2$ and $\gamma_j = \alpha\gamma_{j-1}$, $j \geq 2$ are the covariance functions for the error term.

The variance ratio is, therefore,

$$(B1) \quad V_y = \frac{\sigma_{k,y}^2}{k\sigma_{1,y}^2} = \frac{s\gamma_0 + 2\gamma_1\sum_{j=1}^{s-1}(s-j)\alpha^{j-1}}{s\gamma_0} = 1 + \frac{2\rho_1}{s}\left[\sum_{j=1}^{s-1}(s-j)\alpha^{j-1}\right] = 1 + \frac{2\rho_1}{s}\left[\frac{s(1-\alpha) - (1-\alpha^s)}{(1-\alpha)^2}\right],$$

where $\rho_1$, the first-order autocorrelation of the error term, is given by $\rho_1 = \frac{\gamma_1}{\gamma_0} = \frac{(1+\alpha\theta)(\alpha+\theta)}{1+\theta^2+2\alpha\theta}$, and the last equality in the above expression can be shown to hold as follows:

Let $A \equiv \sum_{j=1}^{s-1}(s-j)\alpha^{j-1}$, then

$$A - \alpha A = \left[\sum_{j=1}^{s-1}(s-j)\alpha^{j-1}\right] - \left[\sum_{j=1}^{s-1}(s-j)\alpha^j\right] = (s-1) - \left[\alpha + \alpha^2 + \cdots + \alpha^{s-2} + \alpha^{s-1}\right] = (s-1) - \bar{A},$$

where $\bar{A} = \sum_{i=1}^{s-1}\alpha^i$. Then, $\bar{A} - \alpha\bar{A} = \left[\sum_{i=1}^{s-1}\alpha^i\right] - \left[\sum_{i=1}^{s-1}\alpha^{i+1}\right] = \alpha - \alpha^s$ which implies that $\bar{A} = \frac{\alpha - \alpha^s}{1-\alpha}$.

Therefore, $A - \alpha A = (s-1) - \left(\dfrac{\alpha - \alpha^s}{1-\alpha}\right)$, which, when solved for $A$, yields $A = \dfrac{s(1-\alpha) - (1-\alpha^s)}{(1-\alpha)^2}$.

\*\*\*\*\*\*\*\*\*\*\*\*\*\*\*\*\*\*\*\*\*\*\*\*\*\*\*\*\*\*\*\*\*\*\*\*\*\*\*\*\*\*\*\*\*\*\*\*\*\*\*\*\*\*\*\*\*\*\*\*\*\*\*\*

**B-II. Evaluating the Variance Ratio for Nonstationary Process with ARMA(1,1) Errors**

What can be said about the values that the variance ratio (B1) can take in this case? While the expression can not be theoretically evaluated, we can establish some facts about it.

**Lemma:** $\dfrac{s(1-\alpha) - (1-\alpha^s)}{(1-\alpha)^2} > 0$ for $|\alpha| < 1$ and $s > 1$.

**Proof:** Using induction we show that (i) the above ratio is positive when $s = 2$ and (ii) if the ratio is positive for a given value $s$, then it is also positive for $s+1$. Therefore, the expression is positive $\forall s \geq 2$. These two statements are shown to be true below.

(i) If $s = 2$, then $A = \dfrac{s(1-\alpha) - (1-\alpha^s)}{(1-\alpha)^2} = 1 > 0$.

(ii) Assume that the ratio is positive for $s$. We need to show that

$$\dfrac{(s+1)(1-\alpha) - (1-\alpha^{(s+1)})}{(1-\alpha)^2} > 0. \text{ But}$$

$$\dfrac{(s+1)(1-\alpha) - (1-\alpha^{s+1})}{(1-\alpha)^2} = \dfrac{s(1-\alpha) - (1-\alpha^s) + \alpha^{s+1} + \alpha^s + 1 - \alpha}{(1-\alpha)^2} = A + \dfrac{1-\alpha^s}{1-\alpha} > 0,$$

where the last inequality holds because $A > 0$ by assumption, and since $|\alpha| < 1$, we have $1 - \alpha^s > 0$ and $1 - \alpha > 0$, regardless of the sign of $\alpha$; therefore, $\dfrac{1-\alpha^s}{1-\alpha} > 0$   **QED**

**Corollary:** It follows from the above lemma and the variance ratio expression given by (B1) in Appendix B-I, that $V_y > 1$ if $\rho_1 > 0$—which is the case for many economic series, $V_y < 1$ if $\rho_1 < 0$, and $V_y = 1$ if $\rho_1 = 0$.

**Proof:** Since $V_y = 1 + \dfrac{2\rho_1}{s} A$ and $A > 0$ as shown by the previous Lemma, we can state that $V_y = 1 + a\ positive\ quantity$, so it is greater than 1. This is similar to the results we obtained for the nonstationary series with AR(1) errors. On the other hand, if $\rho_1 < 0$, then $V_y = 1 + a\ negative\ quantity$

, so it is less than 1. Clearly, when $\rho_1 = 0$, the second term in the variance ratio drops out and it equals 1. This is what we obtained earlier for a nonstationary series with i.i.d. errors.

The following examples further identify variance ratio values conditional on parametric assumptions. Consider $V_y = 1 + \frac{2}{s}\left[\frac{(1+\alpha\theta)(\alpha+\theta)}{1+\theta^2+2\alpha\theta}\right]\left[\frac{s(1-\alpha)-(1-\alpha^s)}{(1-\alpha)^2}\right]$ and consider the following cases.

<u>Example 1</u>: If $\alpha = -\theta$, then (42) implies that $\rho_1 = \frac{(1+\alpha\theta)(\alpha+\theta)}{1+\theta^2+2\alpha\theta} = 0$, and thus $V_y = 1$.

<u>Example 2</u>: If $\alpha = 0$, then the process is an MA(1) and $\rho_1 = \frac{\theta}{1+\theta^2}$. Therefore, the variance ratio will be $V_y = 1 + 2\left(\frac{\theta}{1+\theta^2}\right)\left(\frac{s-1}{s}\right)$. But because $s > 1$, we have $\frac{s-1}{s} > 0$, and since $|\theta| < 1$, the variance ratio satisfies $0 < V_y < 2$ and, therefore, the size of $V_y$ depends on the sign and magnitude of $\theta$. In particular, $V_y > 1$ if $\theta > 0$, $V_y = 1$ if $\theta = 0$ (which is the case of a random walk), and $V_y < 1$ if $\theta < 0$

<u>Example 3</u>: In a more general case with $\alpha \neq 0$, $\theta \neq 0$, and setting $s = 2$, we obtain

$$V_y = 1 + \left[\frac{(1+\alpha\theta)(\alpha+\theta)}{1+\theta^2+2\alpha\theta}\right]\left[\frac{2(1-\alpha)-(1-\alpha^2)}{(1-\alpha)^2}\right] = 1 + \left[\frac{(1+\alpha\theta)(\alpha+\theta)}{1+\theta^2+2\alpha\theta}\right].$$

Here, if (i) $\alpha > 0$ and $\theta > 0$, then $V_y > 1$, (ii) if $\alpha < 0$ and $\theta < 0$, then $V_y < 1$, (iii) if $\alpha = 0$ and $\theta = 0$, then $V_y = 1$, and (iv) if $\alpha\theta < 0$, that is if $\alpha$ and $\theta$ have opposite signs, then the size of the variance ratio depends on their relative values.

****************************************

## B-III. Variance Ratio for Interpolated Nonstationary Process with ARMA(1,1) Errors

Consider the series $x_{i,t}$ which is a linear interpolation of the series $y_{i,t}$ as discussed before. The one-period difference for the interpolated series is $x_{t,i} - x_{t,i-1} = \frac{1}{s}(y_{t,s} - y_{t-1,s})$. The variance of this one-period difference series is given by

(B2) $\quad \sigma_{1,x}^2 = \left(\frac{1}{s}\right)^2 \text{var}(y_{t,s} - y_{t-1,s}) = \left(\frac{1}{s}\right)^2 \sigma_{k,y}^2 = \left(\frac{1}{s}\right)^2 \left[s\gamma_0 + 2\gamma_1 \sum_{j=1}^{s-1}(s-j)\alpha^{j-1}\right].$

The *s*-period, long, variance can be obtained using the facts that

$$x_{t,i} = \frac{i}{s}\left(s\mu + y_{t-1,s} + \sum_{j=1}^{s} u_{t,j}\right) + \frac{s-i}{s} y_{t-1,s} = i\mu + \left(s\mu + y_{t-2,s} + \sum_{j=1}^{s} u_{t-1,j}\right) + \frac{i}{s}\left(\sum_{j=1}^{s} u_{t,j}\right), \text{ and}$$

$$x_{t-1,i} = i\mu + y_{t-2,s} + \frac{i}{s}\left(\sum_{j=1}^{s} u_{t-1,j}\right), \text{ and therefore, } x_{t,i} - x_{t-1,i} = s\mu + \frac{s-i}{s}\left(\sum_{j=1}^{s} u_{t-1,j}\right) + \frac{i}{s}\left(\sum_{j=1}^{s} u_{t,j}\right).$$

The long variance of the interpolated series is therefore given by

(B3)
$$\begin{aligned}
\sigma_{k,x}^2 &= \mathrm{var}\left[s\mu + \frac{s-i}{s}\left(\sum_{j=1}^{s} u_{t-1,j}\right) + \frac{i}{s}\left(\sum_{j=1}^{s} u_{t,j}\right)\right] \\
&= \left(\frac{s-i}{s}\right)^2\left[s\gamma_0 + 2\gamma_1\sum_{j=1}^{s-1}(s-j)\alpha^{j-1}\right] + \left(\frac{i}{s}\right)^2\left[s\gamma_0 + 2\gamma_1\sum_{j=1}^{s-1}(s-j)\alpha^{j-1}\right] \\
&\quad + 2\left(\frac{s-i}{s}\right)\left(\frac{i}{s}\right)\left[\sum_{j=1}^{s-1}(s-j)\gamma_{s-j} + \sum_{j=0}^{s-1}(s-j)\gamma_{s+j}\right] \\
&= \left[\frac{(s-i)^2 + i^2}{s}\right]\left[s\gamma_0 + 2\gamma_1\sum_{j=1}^{s-1}(s-j)\alpha^{j-1}\right] + 2\left(\frac{s-i}{s}\right)\left(\frac{i}{s}\right)\left[\sum_{j=1}^{s-1}(s-j)\gamma_{s-j} + \sum_{j=0}^{s-1}(s-j)\gamma_{s+j}\right].
\end{aligned}$$

In obtaining the above result, we rely on the following Lemma.

**Lemma:** For the above nonstationary ARMA(1,1) process

$$\mathrm{cov}\left[\left(\sum_{j=1}^{s} u_{t-1,j}\right),\left(\sum_{j=1}^{s} u_{t,j}\right)\right] = \sum_{j=1}^{s-1}(s-j)\gamma_{s-j} + \sum_{j=0}^{s-1}(s-j)\gamma_{s+j}.$$

**Proof:**

$$\begin{aligned}
\mathrm{cov}\left[\left(\sum_{j=1}^{s} u_{t-1,j}\right),\left(\sum_{j=1}^{s} u_{t,j}\right)\right] &= E\left[\left(\sum_{j=1}^{s} u_{t-1,j}\right)\left(\sum_{j=1}^{s} u_{t,j}\right)\right] \\
&= E\left[(u_{t-1,1} + u_{t-1,2} + \cdots + u_{t-1,s-1} + u_{t-1,s})(u_{t,1} + u_{t,2} + \cdots + u_{t,s-1} + u_{t,s})\right] \\
&= E(u_{t-1,1}u_{t,1} + u_{t-1,2}u_{t,1} + \cdots + u_{t-1,s-1}u_{t,1} + u_{t-1,s}u_{t,1}) \\
&\quad + E(u_{t-1,1}u_{t,2} + u_{t-1,2}u_{t,2} + \cdots + u_{t-1,s-1}u_{t,2} + u_{t-1,s}u_{t,2}) \\
&\quad + \\
&\quad \vdots \\
&\quad + E(u_{t-1,1}u_{t,s-1} + u_{t-1,2}u_{t,s-1} + \cdots + u_{t-1,s-1}u_{t,s-1} + u_{t-1,s}u_{t,s-1}) \\
&\quad + E(u_{t-1,1}u_{t,s} + u_{t-1,2}u_{t,s} + \cdots + u_{t-1,s-1}u_{t,s} + u_{t-1,s}u_{t,s})
\end{aligned}$$

$$\begin{aligned}
&= \gamma_1 + \gamma_2 + \gamma_3 + \gamma_4 + \gamma_5 + \cdots + \gamma_{s-1} + \gamma_s \\
&\quad + \gamma_2 + \gamma_3 + \gamma_4 + \gamma_5 + \cdots + \gamma_{s-1} + \gamma_s + \gamma_{s+1} \\
&\quad + \gamma_3 + \gamma_4 + \gamma_5 + \cdots + \gamma_{s-1} + \gamma_s + \gamma_{s+1} + \gamma_{s+2} \\
&\quad \ddots \qquad \ddots \qquad \ddots \qquad \ddots \qquad \ddots \\
&\quad \ddots \qquad \ddots \qquad \ddots \qquad \ddots \qquad \ddots \\
&\quad + \gamma_s + \gamma_{s+1} + \gamma_{s+1} + \gamma_{s+2} + \cdots + \gamma_{2s-2} + \gamma_{2s-1}
\end{aligned}$$

$$\begin{aligned}
&= \gamma_1 + 2\gamma_2 + \cdots + (s-1)\gamma_{s-1} \\
&\quad + s\gamma_s + (s-1)\gamma_{s+1} + \cdots + [s-(s-2)]\gamma_{s+(s-2)} + [s-(s-1)]\gamma_{s+(s-1)}
\end{aligned}$$

$$= \sum_{j=1}^{s-1}(s-j)\gamma_{s-j} + \sum_{j=0}^{s-1}(s-j)\gamma_{s+j}$$

where we used the equalities $E(u_{t-1,1}u_{t,1}) = \gamma_s$, $E(u_{t-1,2}u_{t,1}) = \gamma_{s-1}$, …, $E(u_{t-1,s}u_{t,1}) = \gamma_1$, etc. **QED.**

It is noted that the long variance depends on the periodicity index $i$. To remove the dependency of this expression on $i$, we compute the expected value of the variance estimate, conditional on $i$. Therefore,

$$\sigma^2_{k,x} = \left\{\frac{s^2 - 2s\left[\left(\sum_{i=1}^{s} i\right)/s\right] + 2\left[\left(\sum_{i=1}^{s} i^2\right)/s\right]}{s^2}\right\}\left[s\gamma_0 + 2\gamma_1 \sum_{j=1}^{s-1}(s-j)\alpha^{j-1}\right]$$

$$+ \left\{\frac{2s\left[\left(\sum_{i=1}^{s} i\right)/s\right] - 2\left[\left(\sum_{i=1}^{s} i^2\right)/s\right]}{s^2}\right\}\left[\sum_{j=1}^{s-1}(s-j)\gamma_{s-j} + \sum_{j=0}^{s-1}(s-j)\gamma_{s+j}\right]$$

$$= \left\{\frac{s^2 - s(s+1) + \left[\frac{(s+1)(2s+1)}{3}\right]}{s^2}\right\}\left[s\gamma_0 + 2\gamma_1 \sum_{j=1}^{s-1}(s-j)\alpha^{j-1}\right]$$

$$+ \left\{\frac{s(s+1) - \left[\frac{(s+1)(2s+1)}{3}\right]}{s^2}\right\}\left[\sum_{j=1}^{s-1}(s-j)\gamma_{s-j} + \sum_{j=0}^{s-1}(s-j)\gamma_{s+j}\right]$$

(B3) $$= \left(\frac{2s^2+1}{3s^2}\right)\left[s\gamma_0 + 2\gamma_1 \sum_{j=1}^{s-1}(s-j)\alpha^{j-1}\right] + \left(\frac{s^2-1}{3s^2}\right)\left[\sum_{j=1}^{s-1}(s-j)\gamma_{s-j} + \sum_{j=0}^{s-1}(s-j)\gamma_{s+j}\right]$$

Then, using (B1) and (B3) we can obtain the variance ratio for the long difference ($k = s$) as follows.

$$V_x = \frac{\sigma_{k,x}^2}{k\sigma_{1,x}^2} = \frac{\left(\frac{2s^2+1}{3s^2}\right)\left[s\gamma_0 + 2\gamma_1\sum_{j=1}^{s-1}(s-j)\alpha^{j-1}\right] + \left(\frac{s^2-1}{3s^2}\right)\left[\sum_{j=1}^{s-1}(s-j)\gamma_{s-j} + \sum_{j=0}^{s-1}(s-j)\gamma_{s+j}\right]}{s\left(\frac{1}{s}\right)^2\left[s\gamma_0 + 2\gamma_1\sum_{j=1}^{s-1}(s-j)\alpha^{j-1}\right]}$$

$$= \frac{\left(\frac{2s^2+1}{3s^2}\right)\left[s\gamma_0 + 2\gamma_1\sum_{j=1}^{s-1}(s-j)\alpha^{j-1}\right] + \left(\frac{s^2-1}{3s^2}\right)\gamma_1\left[\sum_{j=1}^{s-1}(s-j)\alpha^{s-j-1} + \sum_{j=0}^{s-1}(s-j)\alpha^{s+j-1}\right]}{s\left(\frac{1}{s}\right)^2\left[s\gamma_0 + 2\gamma_1\sum_{j=1}^{s-1}(s-j)\alpha^{j-1}\right]}$$

(B4)
$$= \left(\frac{2s^2+1}{3s}\right) + \left(\frac{s^2-1}{3s}\right)\gamma_1\left[\frac{\sum_{j=1}^{s-1}(s-j)\alpha^{s-j-1} + \sum_{j=0}^{s-1}(s-j)\alpha^{s+j-1}}{s\gamma_0 + 2\gamma_1\sum_{j=1}^{s-1}(s-j)\alpha^{j-1}}\right]$$

$$= \left(\frac{2s^2+1}{3s}\right) + \left(\frac{s^2-1}{3s}\right)\rho_1\left[\frac{\sum_{j=1}^{s-1}(s-j)\alpha^{s-j-1} + \sum_{j=0}^{s-1}(s-j)\alpha^{s+j-1}}{s + 2\rho_1\sum_{j=1}^{s-1}(s-j)\alpha^{j-1}}\right]$$

where, in deriving the last row, we used the equality $\rho_1 = \frac{\gamma_1}{\gamma_0}$.

(B4) can be further simplified by utilizing simplified expressions for the three arithmetic-geometric sums in (B4):

$$\sum_{j=1}^{s-1}(s-j)\alpha^{s-j-1} = \frac{1-\alpha^{s-1}-(1-\alpha)(s-1)\alpha^{s-1}}{(1-\alpha)^2}$$

$$\sum_{j=0}^{s-1}(s-j)\alpha^{s+j-1} = \frac{(1-\alpha)s\alpha^{s-1} - \alpha^s + \alpha^{2s}}{(1-\alpha)^2}$$

and $\sum_{j=1}^{s-1}(s-j)\alpha^{j-1} = \frac{s(1-\alpha)-(1-\alpha^s)}{(1-\alpha)^2}$

The proof of these equalities is included at the end of this appendix. Plugging the above three into (B4) and simplifying, we obtain

$$V_x = \left(\frac{2s^2+1}{3s}\right) + \left(\frac{s^2-1}{3s}\right)\rho_1 \left[\frac{\sum_{j=1}^{s-1}(s-j)\alpha^{s-j-1} + \sum_{j=0}^{s-1}(s-j)\alpha^{s+j-1}}{s+2\rho_1\sum_{j=1}^{s-1}(s-j)\alpha^{j-1}}\right]$$

$$= \left(\frac{2s^2+1}{3s}\right) + \left(\frac{s^2-1}{3s}\right)\rho_1 \left\{\frac{\left[\frac{1-\alpha^{s-1}-\alpha^{s-1}(1-\alpha)(s-1)}{(1-\alpha)^2}\right] + \left[\frac{s\alpha^{s-1}(1-\alpha)-\alpha^s+\alpha^{2s}}{(1-\alpha)^2}\right]}{s+2\rho_1\left[\frac{s(1-\alpha)-(1-\alpha^s)}{(1-\alpha)^2}\right]}\right\}$$

$$= \left(\frac{2s^2+1}{3s}\right) + \left(\frac{s^2-1}{3s}\right)\rho_1 \left\{\frac{\left[\frac{(1-\alpha^s)^2}{(1-\alpha)^2}\right]}{s+2\rho_1\left[\frac{s(1-\alpha)-(1-\alpha^s)}{(1-\alpha)^2}\right]}\right\}$$

(B5) $$= \left(\frac{2s^2+1}{3s}\right) + \frac{\rho_1(s^2-1)(1-\alpha^s)^2}{3s^2(1-\alpha)^2 + 6s\rho_1\left[s(1-\alpha)-(1-\alpha^s)\right]}$$

Using the expression for $\rho_1$, we obtain the final expression for the variance ratio of the interpolated series:

(B6) $$V_x = \left(\frac{2s^2+1}{3s}\right) + \frac{(1+\alpha\theta)(\alpha+\theta)(s^2-1)(1-\alpha^s)^2}{3s^2(1-\alpha)^2(1+\theta^2+2\alpha\theta) + 6s(1+\alpha\theta)(\alpha+\theta)\left[s(1-\alpha)-(1-\alpha^s)\right]}$$

**Lemmas**

Simplification for the three arithmetic-geometric sums used to obtain (B5) can be proved using the following three lemmas.

**Lemma:** If $s>1$ and $|\alpha|<1$, then $\sum_{j=0}^{s-1}(s-j)\alpha^{s+j-1} = \frac{s\alpha^{s-1}(1-\alpha)-\alpha^s+\alpha^{2s}}{(1-\alpha)^2}$.

**Proof:** Let $B \equiv \sum_{j=0}^{s-1}(s-j)\alpha^{s+j-1}$, and multiply B by $\alpha$, and subtract it from itself to obtain

$$B-\alpha B = \left[\sum_{j=0}^{s-1}(s-j)\alpha^{s+j-1}\right] - \alpha\left[\sum_{j=0}^{s-1}(s-j)\alpha^{s+j-1}\right] = s\alpha^{s-1} - \left[\alpha^s + \alpha^{s+1} + \cdots + \alpha^{2s-2} + \alpha^{2s-1}\right].$$

But the bracketed expression equals $\frac{\alpha^s - \alpha^{2s}}{1-\alpha}$, which upon substitution yields

$$B - \alpha B = s\alpha^{s-1} - \left(\frac{\alpha^s - \alpha^{2s}}{1-\alpha}\right) \quad \text{or} \quad B = \frac{s\alpha^{s-1}(1-\alpha) - \alpha^s + \alpha^{2s}}{(1-\alpha)^2}. \qquad \textbf{QED}$$

**Lemma:** If $s > 1$ and $|\alpha| < 1$, then $\sum_{j=1}^{s-1}(s-j)\alpha^{s-j-1} = \dfrac{1-\alpha^{s-1} - \alpha^{s-1}(1-\alpha)(s-1)}{(1-\alpha)^2}$.

**Proof:** Let $C \equiv \sum_{j=1}^{s-1}(s-j)\alpha^{s-j-1}$ and multiply C by $\alpha$, and subtract it from itself to obtain

$$C - \alpha C = \left[\sum_{j=1}^{s-1}(s-j)\alpha^{s-j-1}\right] - \alpha\left[\sum_{j=1}^{s-1}(s-j)\alpha^{s-j-1}\right] = \left[1 + \alpha^1 + \alpha^2 + \cdots + \alpha^{s-3} + \alpha^{s-2}\right] - (s-1)\alpha^{s-1}.$$

The bracketed term equals $\dfrac{1-\alpha^{s-1}}{1-\alpha}$, which upon substitution and simplification results in the desired expression; e.g., $C - \alpha C = \dfrac{1-\alpha^{s-1}}{1-\alpha} - (s-1)\alpha^{s-1}$ so, $C = \dfrac{1-\alpha^{s-1} - \alpha^{s-1}(1-\alpha)(s-1)}{(1-\alpha)^2}$. **QED**

**Lemma:** If $s > 1$ and $|\alpha| < 1$, then $\sum_{j=1}^{s-1}(s-j)\alpha^{j-1} = \dfrac{s(1-\alpha) - (1-\alpha^s)}{(1-\alpha)^2}$.

**Proof:** Let $D \equiv \sum_{j=1}^{s-1}(s-j)\alpha^{j-1}$ and, in a manner similar to the previous lemmas, multiply D by $\alpha$, and subtract it from itself to obtain

$$D - \alpha D = \left[\sum_{j=1}^{s-1}(s-j)\alpha^{j-1}\right] - \alpha\left[\sum_{j=1}^{s-1}(s-j)\alpha^{j-1}\right] = (s-1) - \left[\alpha^1 + \alpha^2 + \cdots + \alpha^{s-2} + \alpha^{s-1}\right].$$

The bracketed term equals $\dfrac{\alpha - \alpha^s}{1-\alpha}$, which upon substitution and simplification results in the desired expression; e.g., $D - \alpha D = (s-1) - \left(\dfrac{\alpha - \alpha^s}{1-\alpha}\right)$ so $D = \dfrac{s(1-\alpha) - (1-\alpha^s)}{(1-\alpha)^2}$. **QED**